\documentclass[sigconf, nonacm]{acmart}

\newcommand\vldbdoi{XX.XX/XXX.XX}
\newcommand\vldbpages{XXX-XXX}
\newcommand\vldbvolume{14}
\newcommand\vldbissue{1}
\newcommand\vldbyear{2020}
\newcommand\vldbauthors{\authors}
\newcommand\vldbtitle{\shorttitle} 
\newcommand\vldbavailabilityurl{https://anonymous.4open.science/r/TIMEST/}
\newcommand\vldbpagestyle{plain} 

\usepackage{algorithm}
\usepackage{array}
\usepackage[noend]{algpseudocode}
\usepackage{balance}
\usepackage{multirow}
\usepackage{paralist}
\usepackage{subcaption}
\usepackage{colortbl}
\usepackage{enumitem}
\usepackage[export]{adjustbox}
\usepackage[compact]{titlesec}
 \titlespacing*{\section}{0pt}{3pt}{3pt}
 \titlespacing*{\subsection}{0pt}{1pt}{1pt}

\colorlet{Green1}{green!25}
\definecolor{Green2}{HTML}{99d8c9}

\newcommand{\THISWORK}{{\fontfamily{lmss}\selectfont 
TIMEST}}

\newcommand{\Revision}[1]{{\color{black!100}{#1}}}
\newtheorem{theorem}{Theorem}[section]
\newtheorem{claim}[theorem]{Claim}

\newtheorem{definition}[theorem]{Definition}

\newcommand{\pluseq}{\mathrel{+}=}

\newcommand{\tempout}[3]{d^+_#1[#2,#3]}
\newcommand{\tempin}[3]{d^-_#1[#2,#3]}

\newcommand{\outlist}[3]{\Lambda^+_#1[#2, #3]}
\newcommand{\inlist}[3]{\Lambda^-_#1[#2, #3]}
\newcommand{\alphalist}[5]{\Lambda^{#1}_#2[#3, (#4, #5)]}

\newcommand{\multiedgelist}[5]{El^{#1}_#2[#3, (#4, #5)]}

\newcommand{\Sec}[1]{\hyperref[sec:#1]{Section~\ref*{sec:#1}}} 
\newcommand{\Eqn}[1]{\hyperref[eq:#1]{(\ref*{eq:#1})}} 
\newcommand{\Fig}[1]{\hyperref[fig:#1]{Figure\,\ref*{fig:#1}}} 
\newcommand{\Tab}[1]{\hyperref[tab:#1]{Table\,\ref*{tab:#1}}} 
\newcommand{\Thm}[1]{\hyperref[thm:#1]{Theorem\,\ref*{thm:#1}}} 
\newcommand{\Fact}[1]{\hyperref[fact:#1]{Fact\,\ref*{fact:#1}}} 
\newcommand{\Lem}[1]{\hyperref[lem:#1]{Lemma\,\ref*{lem:#1}}} 
\newcommand{\Prop}[1]{\hyperref[prop:#1]{Prop.~\ref*{prop:#1}}} 
\newcommand{\Cor}[1]{\hyperref[cor:#1]{Corollary~\ref*{cor:#1}}} 
\newcommand{\Conj}[1]{\hyperref[conj:#1]{Conjecture~\ref*{conj:#1}}} 
\newcommand{\Def}[1]{\hyperref[def:#1]{Definition~\ref*{def:#1}}} 
\newcommand{\Alg}[1]{\hyperref[algo:#1]{Algorithm~\ref*{algo:#1}}} 
\newcommand{\Ex}[1]{\hyperref[ex:#1]{Example.~\ref*{ex:#1}}} 
\newcommand{\Clm}[1]{\hyperref[clm:#1]{Claim~\ref*{clm:#1}}} 
\newcommand{\Step}[1]{\hyperref[step:#1]{Step~\ref*{step:#1}}} 
\newcommand{\EX}{\mathbf{E}}

\usepackage{tikz}
\tikzset{%
    baseline,
    inner sep=2pt,
    minimum height=12pt,
    rounded corners=2pt  
}
\newcommand{\code}[1]{\mbox{
    \ttfamily
    \tikz \node[anchor=base,fill=black!12]{#1};
}}

\AtBeginDocument{%
  \providecommand\BibTeX{{%
    \normalfont B\kern-0.5em{\scshape i\kern-0.25em b}\kern-0.8em\TeX}}}

\setcopyright{acmlicensed}
\copyrightyear{2024}
\acmYear{2024}
\acmDOI{XXXXXXX.XXXXXXX}

\acmConference[Conference acronym 'XX]{Make sure to enter the correct
  conference title from your rights confirmation emai}{June 03--05,
  2024}{Woodstock, NY}
\acmISBN{978-1-4503-XXXX-X/18/06}

\begin{document}

\newcolumntype{P}[1]{>{\centering\arraybackslash}p{#1}}
\newcolumntype{C}[1]{>{\centering\arraybackslash}m{#1}}   
\newcolumntype{R}[1]{>{\raggedleft\arraybackslash}m{#1}}  
\newcolumntype{L}[1]{>{\raggedright\arraybackslash}m{#1}}  

\colorlet{Green1}{green!25}
\definecolor{Green2}{HTML}{99d8c9}

\title{TIMEST: \underline{T}emporal \underline{I}nformation \underline{M}otif \underline{E}stimator Using \underline{S}ampling \underline{T}rees}


\author{Yunjie Pan}
\affiliation{%
  \institution{University of Michigan}
  \city{Ann Arbor}
  \state{Michigan}
  \country{USA}
}
\email{panyj@umich.edu}

\author{Omkar Bhalerao}
\affiliation{%
  \institution{University of California, Santa Cruz}
  \city{Santa Cruz}
  \state{California}
  \country{USA}
}
\email{obhalera@ucsc.edu}

\author{C. Seshadhri}
\affiliation{%
  \institution{University of California, Santa Cruz}
  \city{Santa Cruz}
  \state{California}
  \country{USA}
}
\email{sesh@ucsc.edu}

\author{Nishil Talati}
\affiliation{%
  \institution{University of Michigan}
  \city{Ann Arbor}
  \state{Michigan}
  \country{USA}
}
\email{talatin@umich.edu}

\renewcommand{\shortauthors}{Trovato and Tobin, et al.}

\begin{abstract}

The mining of pattern subgraphs, known as motifs, is a core task in the field of graph mining. 
Edges in real-world networks often have timestamps, so there is a need for \emph{temporal motif mining}. A temporal motif is a richer structure that imposes timing constraints on the edges of the motif.
Temporal motifs have used to analyze social networks, financial transactions, and biological networks.

Motif counting in temporal graphs is particularly challenging. A graph with millions of edges can have trillions of temporal motifs, since the same edge can occur with multiple timestamps. There is a combinatorial explosion of possibilities, and state-of-the-art algorithms cannot manage motifs with more than four vertices.

In this work, we present \THISWORK: a general, fast, and accurate estimation algorithm to count temporal motifs of arbitrary sizes in temporal networks.
Our approach introduces a temporal spanning tree sampler that leverages weighted sampling to generate substructures of target temporal motifs. This method carefully takes a subset of temporal constraints of the motif that can be jointly and efficiently sampled.
\THISWORK\ uses randomized estimation techniques to obtain accurate estimates of motif counts.

We give theoretical guarantees on the running time and approximation guarantees of \THISWORK. We perform an extensive experimental evaluation and show that \THISWORK\ is both faster \emph{and} more accurate than previous algorithms.
Our CPU implementation exhibits an average speedup of 28$\times$ over state-of-the-art GPU implementation of the exact algorithm,  and 6$\times$ speedup over SOTA approximate algorithms while consistently showcasing less than $5\%$ error in most cases. For example, \THISWORK\ can count the number of instances of a financial fraud temporal motif
in four minutes with $0.6\%$ error, while exact methods take
more than \emph{two days}.

\end{abstract}

\maketitle

\pagestyle{\vldbpagestyle}
\begingroup\small\noindent\raggedright\textbf{PVLDB Reference Format:}\\
\vldbauthors. \vldbtitle. PVLDB, \vldbvolume(\vldbissue): \vldbpages, \vldbyear.\\
\endgroup
\begingroup
\renewcommand\thefootnote{}\footnote{\noindent
This work is licensed under the Creative Commons BY-NC-ND 4.0 International License. Visit \url{https://creativecommons.org/licenses/by-nc-nd/4.0/} to view a copy of this license. For any use beyond those covered by this license, obtain permission by emailing \href{mailto:info@vldb.org}{info@vldb.org}. Copyright is held by the owner/author(s). Publication rights licensed to the VLDB Endowment. \\
\raggedright Proceedings of the VLDB Endowment, Vol. \vldbvolume, No. \vldbissue\ %
ISSN 2150-8097. \\
\href{https://doi.org/\vldbdoi}{doi:\vldbdoi} \\
}\addtocounter{footnote}{-1}\endgroup

\ifdefempty{\vldbavailabilityurl}{}{
\vspace{.3cm}
\begingroup\small\noindent\raggedright\textbf{PVLDB Artifact Availability:}\\
The source code, data, and/or other artifacts have been made available at \url{\vldbavailabilityurl}.
\endgroup
}

\section{Introduction}

\begin{figure}[t]
    \centering
    \begin{subfigure}{0.48\columnwidth}
        \includegraphics[width=\textwidth]{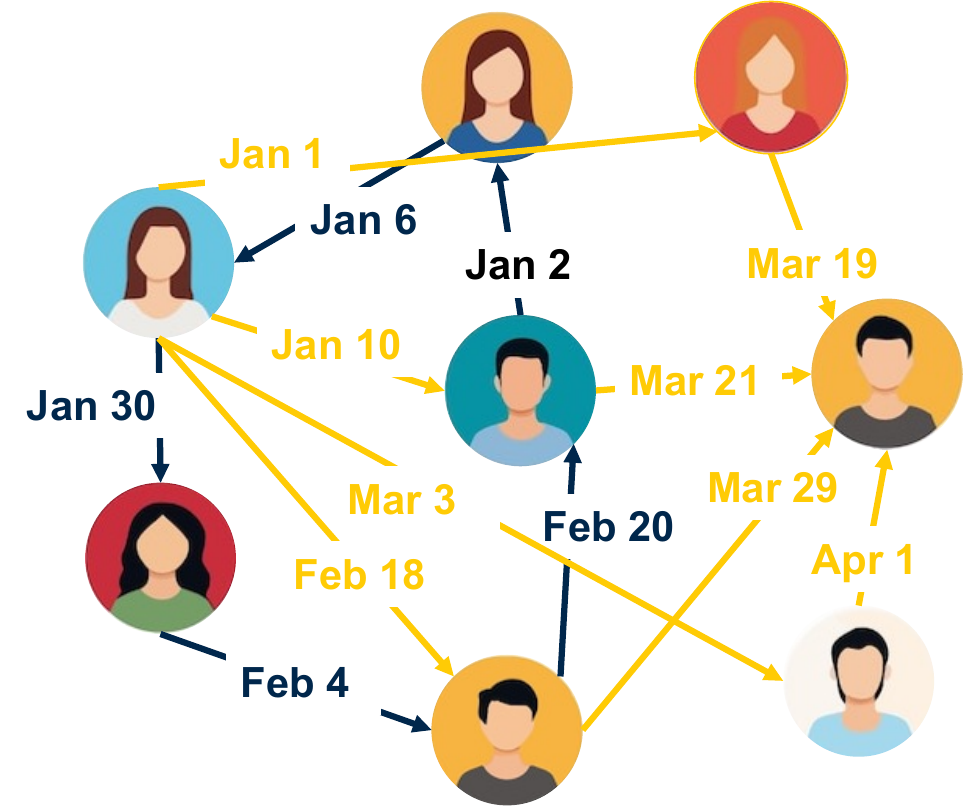}
        \caption{Financial transactions graph}
        \label{fig:money-laundering-example}
    \end{subfigure}
    \hfill
    \begin{subfigure}{0.48\columnwidth}
    \begin{subfigure}{0.47\columnwidth}
    \includegraphics[width=\linewidth]{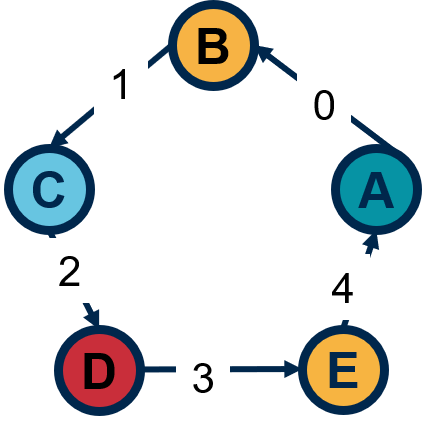}
    \caption{5-node cycle}
    \label{fig:money-laundering-5-cycle}
    \end{subfigure} 
    \begin{subfigure}{0.48\columnwidth}
    \includegraphics[width=\linewidth]{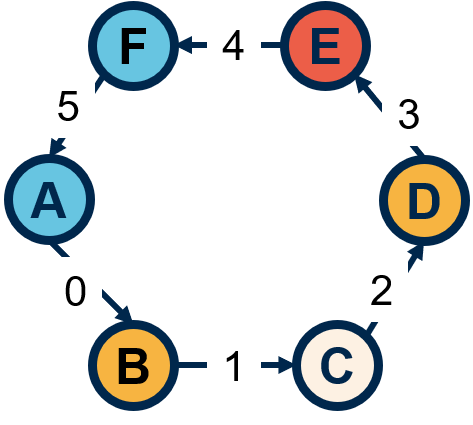}
        \caption{6-node cycle}
        \label{fig:money-laundering-6-cycle}
    \end{subfigure}
    \\
    \begin{subfigure}{0.53\columnwidth}
    \includegraphics[width=\linewidth]{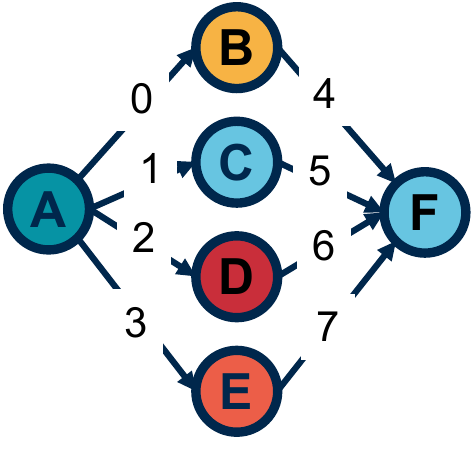}
        \caption{scatter-gather}
        \label{fig:money-laundering-scatter-gather}
    \end{subfigure}
    \begin{subfigure}{0.43\columnwidth}
    \includegraphics[width=\linewidth]{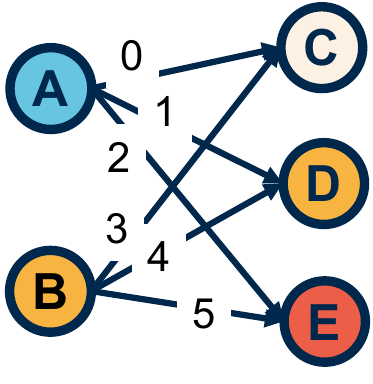}
        \caption{bipartite}
        \label{fig:money-laundering-bipartite}
    \end{subfigure}
    \end{subfigure}
    
    \caption{(a) An example of financial transactions with timestamps. Different money-laundering and gambling pattens~\cite{altman2024realistic,liu2023temporal} including (b) 5-node simple-cycle; 
    (c) 6-node simple-cycle; (d) scatter-gather; and (e) bipartite.}
    \label{fig:money-laundering}
\end{figure}

A central tool in the analysis of large networks is \emph{motif mining}~\cite{mining2006data, milo2002network,paranjape2017motifs,suzumura2021anti,liu2023temporal}. 
A motif is a small pattern graph that indicates some special structure in a larger input graph. 
Motifs offer valuable insight into the graph structure,
and are analogous to queries that detail a set of edge constraints.
The algorithmic problem
of motif mining, often called subgraph counting/finding in the algorithms literature, has
a rich history in the data mining literature~\cite{shen2002network, alon2007network, newman2003structure}. (See survey~\cite{seshadhri2019scalable}.)

Most real-world graphs are \emph{temporal}, where edges come with timestamps. There
is a surge of recent work on mining temporal graphs,
 especially searching for temporal motifs (refer to survey~\cite{GoIeSa-tut}). These works show temporal motifs can capture richer information
than standard motifs~\cite{kovanen2011temporal, pan2011path}.
Temporal motifs have been used for user behavior characterization on social networks~\cite{kovanen2013temporal, lahiri2007structure, paranjape2017motifs}, detecting financial fraud~\cite{hajdu2020temporal,liu2023temporal}, characterizing function of biological networks~\cite{pan2011path,hulovatyy2015exploring}, and improving graph neural networks~\cite{bouritsas2022improving, liu2023temporal}.

A temporal graph can be viewed as a graph database with a rich set of attributes such as vertex/node labels and edge timestamps. 
Mining motifs in temporal graphs is analogous to executing queries in traditional database systems.

\begin{figure}[t]
    \centering
    \begin{subfigure}{\linewidth}
        \includegraphics[width=0.95\linewidth]{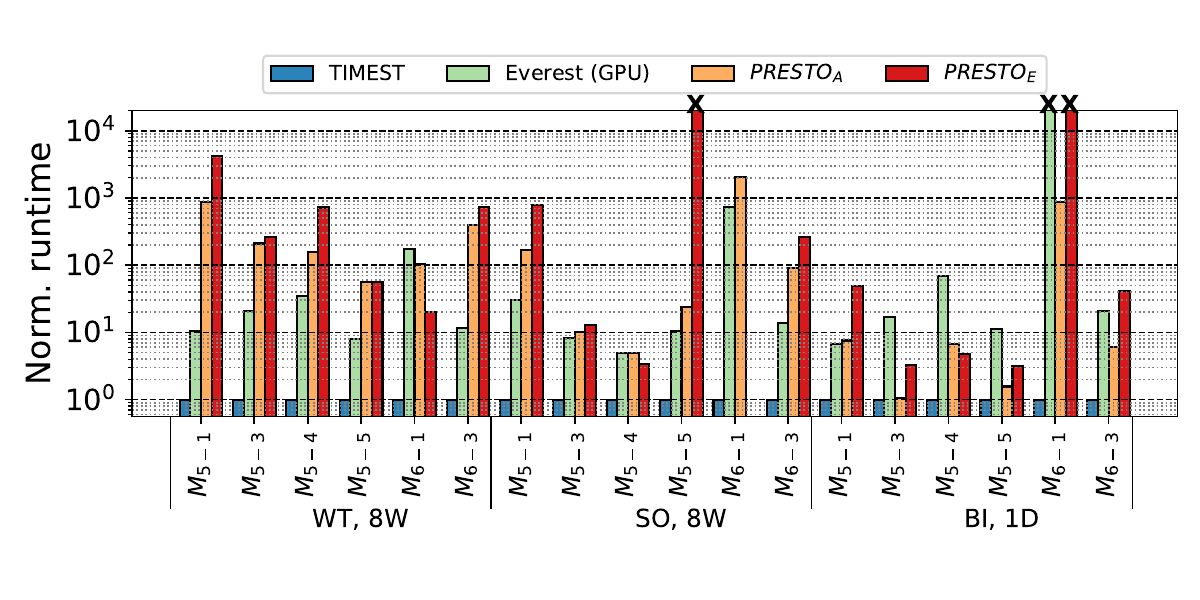}
        \vspace{-5mm}
        \caption{Runtime of prior works normalized to \THISWORK. \THISWORK\ has an average speedup of 28$\times$ compared to previous works.}
        \label{fig:runtime}
    \end{subfigure}
    \begin{subfigure}{\linewidth}
        \centering
        \includegraphics[width=0.95\linewidth]{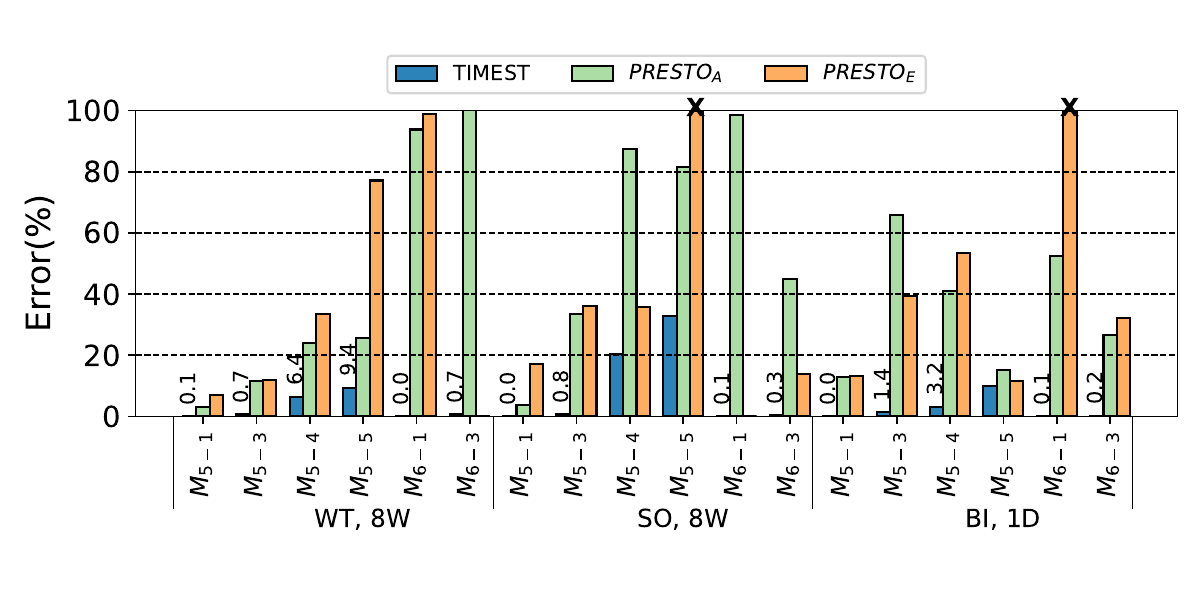}
        \vspace{-5mm}
        \caption{Relative error (\%) of approximate algorithms. \THISWORK\ is almost always the most accurate while always the fastest.}
        \label{fig:error}
    \end{subfigure}
    \caption{We show the runtime improvement and estimation errors of \THISWORK\ versus Everest~\cite{yuan2023everest} (GPU) and approximate algorithms~\cite{sarpe2021presto} ($PRESTO_A$ and $PRESTO_E$) on various datasets and motifs. Everest is tested on an NVIDIA A40 GPU, while others are evaluated on a CPU using 32 threads. "x" means run out of memory (OOM) or timeout (> 1 day).}
\end{figure}

\subsection{Problem Definition} \label{sec:problem}

\begin{figure*}

\includegraphics[width=\linewidth]{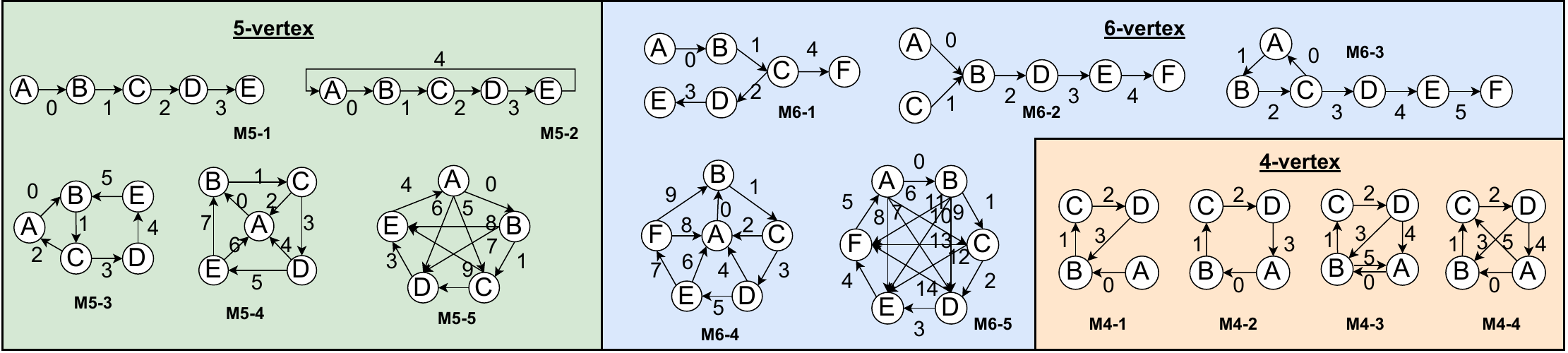}
\caption{\Revision{Connected 4-, 5- and 6-vertex temporal motifs used for evaluation.}}
    \label{fig:temp_motifs}
\end{figure*}
%

We set some notation. The input is 
a multigraph $G = (V(G), E(G))$, with $n$ vertices and $m$ edges. Each edge in $G$ is a tuple $(u,v,t)$ where $u$ and $v$ are source and destination vertices, and $t$ is a positive integer timestamp. \Revision{Following prior work~\cite{mackey2018chronological,yuan2023everest}, we assume that each tuple $(u,v,t)$ can appear only once.} (The same $u, v$ can have multiple edges between them
with different timestamps, and there can be different tuples with the same timestamp.) For a temporal edge $e$, let $t(e)$ to denote its timestamp.
A temporal motif involves a standard graph motif
with a constraint on the time window the motif occurs in, and a constraint on the order of edges.
We formally define temporal motifs and temporal matches, following Paranjape \textit{et al.}~\cite{paranjape2017motifs}.

\begin{definition} \label{def:temp-motif} A \emph{temporal motif} is a triple $M=(H,\pi, \delta)$ where (i) $H = (V(H), E(H))$ is a directed pattern graph, $(ii)$ $\pi$ is \Revision{an ordered list (alternately, a permutation)} of the edges of $H$, and (iii) $\delta$ is a positive integer.

The \Revision{ordered list} $\pi$ specifies the time ordering of edges, and $\delta$ specifies the length of the time interval in which edges must occur.
\end{definition}

\begin{definition} \label{def:temp-match} Consider an input temporal graph $G = (V(G),E(G))$ and a temporal pattern $M = (H,\pi,\delta)$. An \emph{$M$-match} is a pair $(\phi_{V}, \phi_{E})$ 1-1 maps $\phi_V:V(H) \to V(G)$, $\phi_{E}:E(H) \to E(G)$ satisfying the following conditions.
\begin{asparaitem}
\item  (Matching the edges) $\forall (u,v) \in E(H), \phi_{E}(u,v) = (\phi_{V}(u), \phi_{V}(v))$
    \item (Matching the pattern)  $\forall (u,v) \in E(H)$, $(\phi_V(u),\phi_V(v)) \in E(G)$. 
    \item (Edges ordered correctly) The timestamps of the edges in the match follow the ordering $\pi$. Formally, $\forall e, e' \in E(H)$, $\pi(e) < \pi(e')$ iff $t(\phi_E(e)) < t(\phi_E(e'))$.
    \item (Edges in time interval) All edges of the match occur within $\delta$ time range. Formally, $\forall e, e' \in E(H)$, $|t(\phi_E(e)) - t(\phi_E(e'))| \leq \delta$.
\end{asparaitem}
\end{definition}

\Fig{money-laundering-example} shows an example of financial transactions, represented as a graph. We also four temporal motifs, which have been explicitly mentioned as
 indications of money laundering~\cite{suzumura2021anti,shadrooh2024smotef}.
Note that the edges of the cycle must occur in temporal order (given the edge label), for this to represent
a valid cyclic money flow. More temporal motifs are given in \Fig{temp_motifs}.
\Revision{We note that a natural generalization is to impose a \emph{partial order} of time constraints, which we discuss more in future work.}

\subsection{The Challenge} \label{sec:challenge}

Temporal motif counting is particularly difficult because of \emph{combinatorial explosion}.
There can be thousands of temporal edges between the same vertices, which leads
to a exponentially larger search space for motifs. For example, there are trillions of temporal 5-cliques in Bitcoin graph with 100 million edges.  
Exact methods based on enumeration or exploration 
cannot avoid this massive computation~\cite{paranjape2017motifs,mackey2018chronological}.
Many techniques for reducing the search space using graph properties
like the degeneracy and dynamic programming are tailored for simple graphs~\cite{seshadhri2014wedge,pinar2017escape,bressan2018motif,seshadhri2019scalable}. These techniques cannot incorporate ordering constraints on edges.
In general, when the motif has four vertices, no exact
method is able to get results on graphs with 100M edges even in a day with commodity
hardware~\cite{paranjape2017motifs,kumar20182scent,pashanasangi2021faster}.

There are two approaches to temporal motif counting. The usual method
is to design general purpose algorithms that work for (potentially) any motif.
Exact motif counting algorithms rely on explicit subgraph enumeration that suffers a massive computational explosion~\cite{paranjape2017motifs,mackey2018chronological,yuan2023everest}.
There are general purpose estimators for all kinds of temporal motifs, like IS~\cite{liu2019sampling}, ES~\cite{wang2020efficient} and PRESTO~\cite{sarpe2021presto}.
However, they tend to perform poorly on larger motifs, since
they rely on exact algorithm in subsampled intervals.
For example, our evaluation shows that PRESTO runs for more than 5 hours estimating a 5-clique (\textit{i.e.,} $M_{5-5}$ in \Fig{temp_motifs}) with a high 25\% estimation error.


Other methods analyze specific motifs and designs specialized algorithms. 
These include 2SCENT~\cite{kumar20182scent} (for simple cycles), DOTT~\cite{pashanasangi2021faster} (for triangles), Gao \textit{et al.}~\cite{gao2022scalable} (caters to 2-, 3-node, 3-edge motifs), and TEACUPS~\cite{pan2024accurate} (for 4-node motifs), alongside other specialized motifs such as bi-triangle~\cite{yang2021efficient}, bi-clique~\cite{ye2023efficient}, and butterfly~\cite{cai2023efficient, pu2023sampling} motifs for bipartite networks.
While these often work well for the tailored motif, they do not
give general purpose algorithms. Common motifs with more than 5 nodes occurs
in money laundering applications~\cite{shadrooh2024smotef,suzumura2021anti,altman2024realistic}. These motifs are shown in \Fig{money-laundering},
and cannot be handled efficiently by existing methods.

This leads to the main question of our paper: \emph{do there exist general purpose temporal
motif counting algorithms that can scale beyond current motif sizes?}

\begin{figure*}
\includegraphics[width=0.8\linewidth]{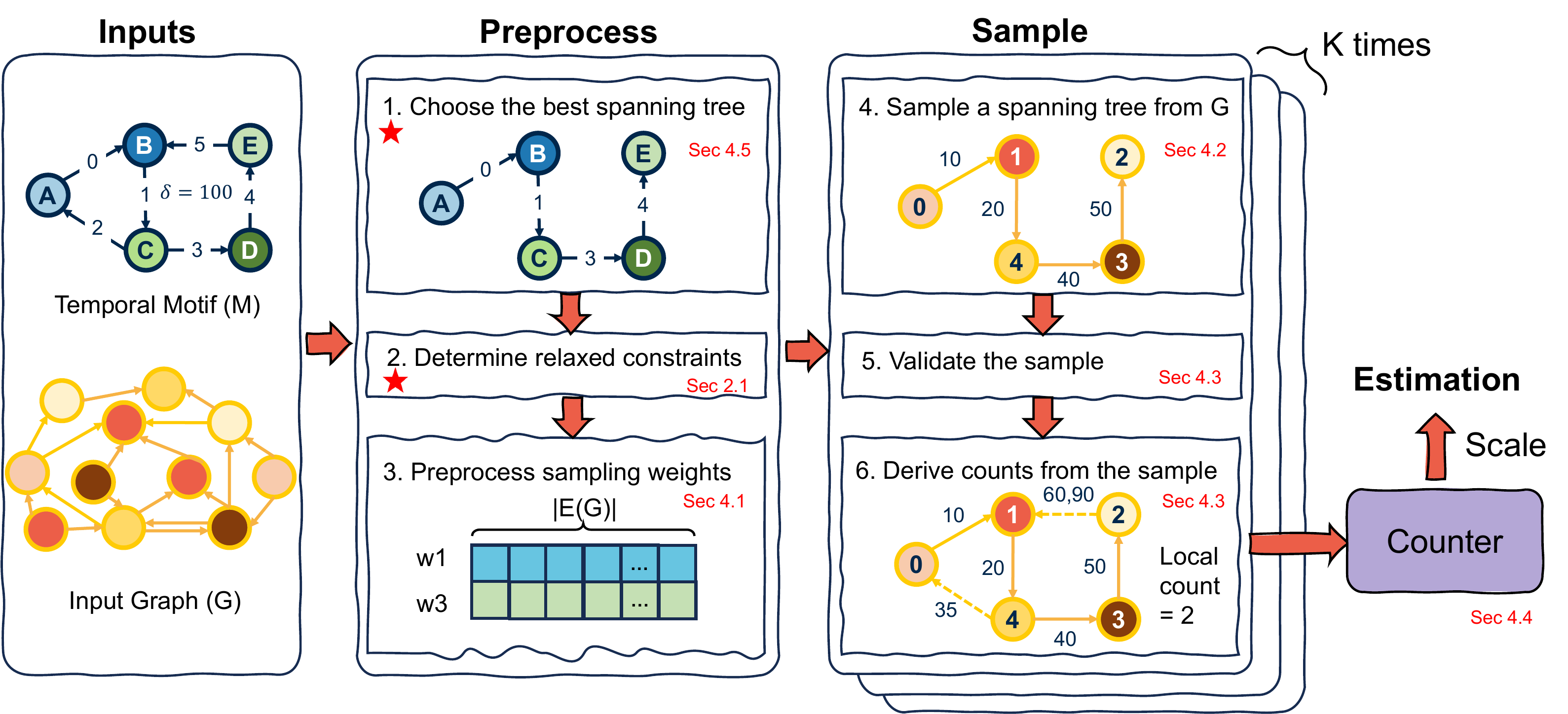}
\vspace{-3mm}
\caption{Overall process of estimating temporal motif $M$ counts in the input graph $G$.}
\label{fig:overall-process}
\end{figure*}

\begin{figure*}
    \captionsetup[subfigure]{justification=centering}
    \centering
    \begin{subfigure}{0.13\linewidth}
                \includegraphics[width=\textwidth]{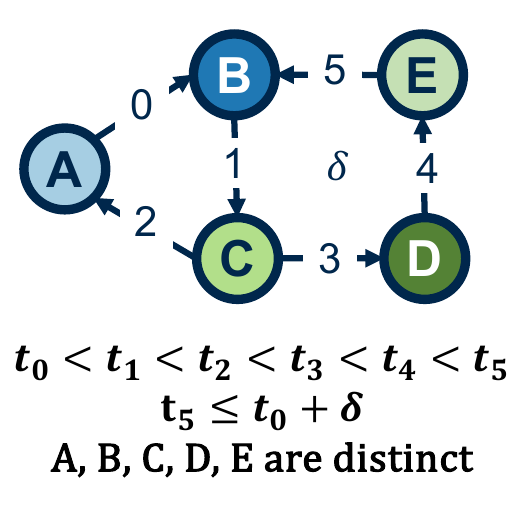} 
                \caption{Original Constraints}
                \label{fig:constraint_all}
    \end{subfigure}
    \begin{subfigure}{0.42\textwidth}
    \includegraphics[width=\textwidth]{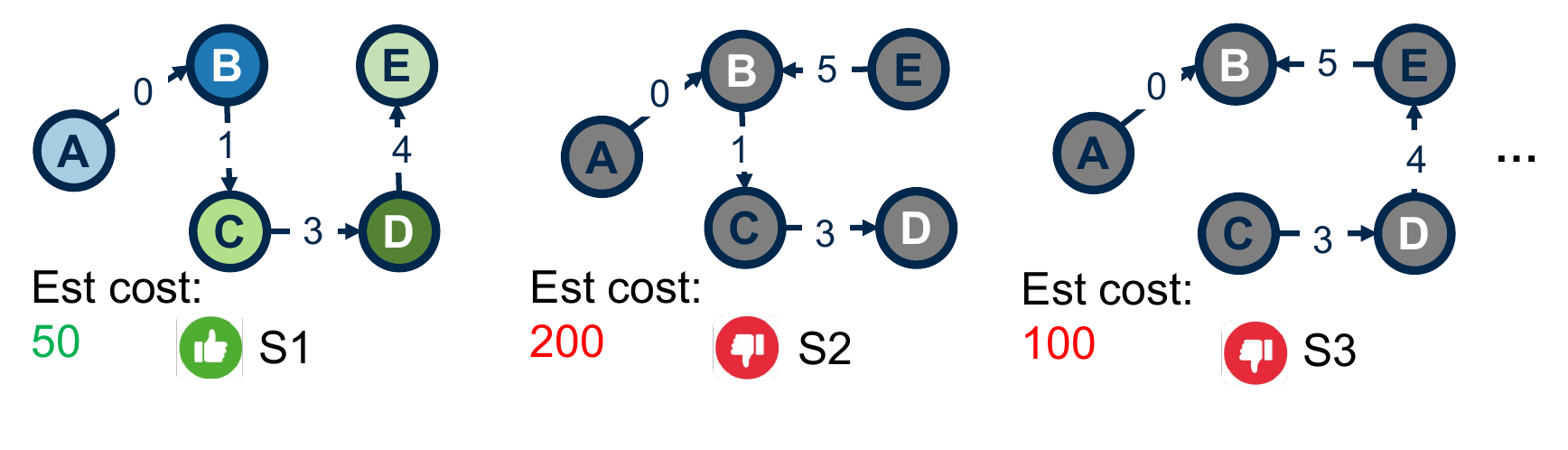}
    \caption{Use heuristics to choose the best spanning tree S of motif M}
    \label{fig:spanning_trees}
    \end{subfigure}
    \begin{subfigure}{0.13\textwidth}
    \includegraphics[width=\textwidth]{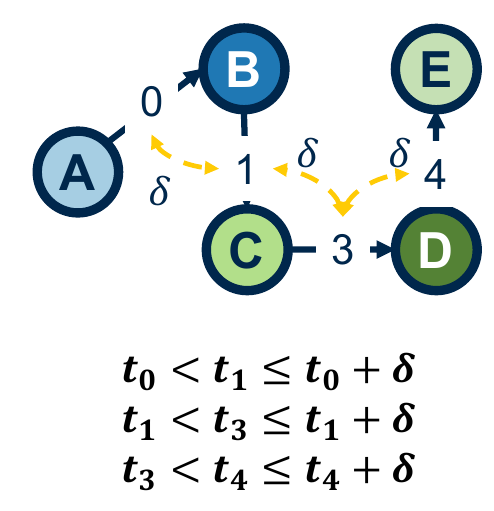} 
    \vspace{-4mm}
    \caption{Constraint 1}
    \label{fig:constraint_1}
    \end{subfigure}
    \begin{subfigure}{0.13\textwidth}
    \includegraphics[width=\textwidth]{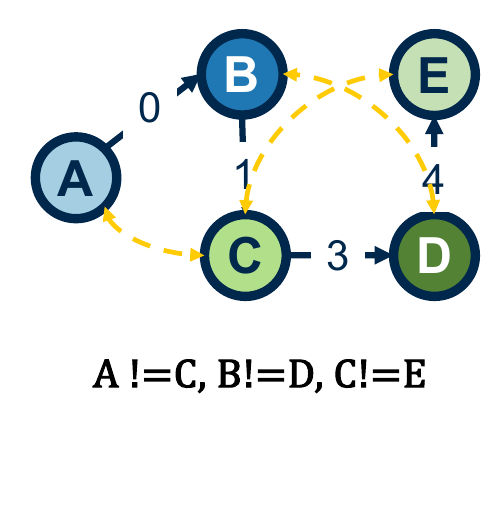} 
    \vspace{-4mm}
    \caption{Constraint 2}
    \label{fig:constraint_2}
    \end{subfigure}
    \begin{subfigure}{0.13\textwidth}
    \includegraphics[width=\textwidth]{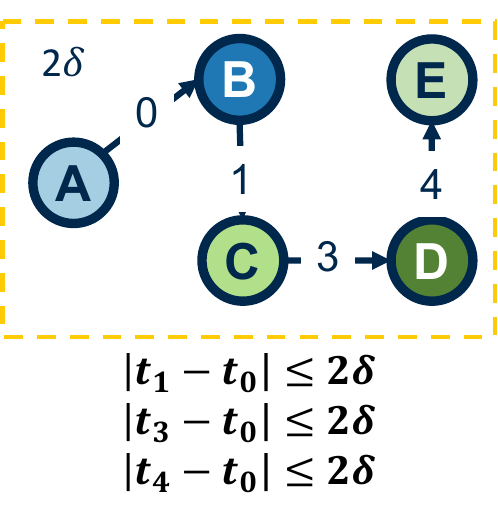} 
    \vspace{-4mm}
    \caption{Constraint 3}
    \label{fig:constraint_3}
    \end{subfigure}
    
    \caption{An example of mining motif $M_\text{5-3}$. (a) The standard approach strictly enforces all constraints for a temporal motif match. (b) Multiple spanning trees of the motif exist, and we use a heuristic to select the optimal one that has the lowest estimated sampling cost (Est cost). We then relax specific constraints and apply three partial constraints to the spanning tree: (c) enforcing edge order and $\delta$-window constraints for adjacent edges; (d) ensuring distinct end vertices for adjacent edges; and (e) constraining all edges to fall within a $2\delta$ time window.}
    \label{fig:constraints}
\end{figure*}

\subsection{Our Results} \label{sec:results}

Our main result is \THISWORK, a general and efficient
randomized (approximate) temporal motif mining algorithm. 
\THISWORK\ is a general purpose method that works for any motif.
As a concrete example, it is able to efficiently approximate counts for all
the example motifs in \Fig{temp_motifs}.
We introduce a new technique of \emph{temporal spanning tree
sampling} \Revision{that generalizes previous temporal results~\cite{PaBh+24}
and older non-temporal methods~\cite{seshadhri2014wedge,jha2015path,wang2017moss}}. We list out some specific contributions
of our work.

{\bf Temporal spanning tree sampling:} Our primary contribution
is a new technique of \emph{temporal spanning tree
sampling}, that generalizes a recent method of temporal
path sampling~\cite{PaBh+24}. To count temporal motifs of arbitrary size,
we find an appropriate spanning tree of the motif.
These spanning trees satisfy partial temporal constraints
of this motif. We then sample many such trees, and estimate
the fraction of them that ``extend" to the motif.
 
A direct temporal spanning tree
would simply take a subset of the edges \emph{and} the corresponding
temporal constraints. But it is not clear how to efficiently
sample such a tree with these constraints. We show
that by a careful relaxation of some of the temporal constraints,
an efficient sampler can be constructed. This sampler quickly
preprocesses the graph and can output any desired number of uniform
random spanning trees.

%
%
%

{\bf Sample reduction techniques:} To reduce the sample 
complexity, we design heuristics that choose the ``best" temporal
spanning tree. We can formally show that the temporal spanning tree
with the fewest matches is the best choice, but finding
this tree would itself require a large number of motif counts. We design 
heuristics to narrow down on the best spanning tree, which leads to 
lower error with the same sampling budget.

{\bf Practical efficiency of \THISWORK:} We perform a comprehensive evaluation across various datasets (\Tab{dataset}) and a large collection of complex motifs (\Fig{temp_motifs}), up to $6$ vertices. 
We compare with a state-of-the-art
exact algorithm that uses GPUs (Everest~\cite{yuan2023everest}) and versions of the best randomized method
(Presto~\cite{sarpe2021presto}). Results are in \Fig{runtime}. \THISWORK\ is \emph{always} faster, with a typical 10x-100x speedup
over both existing methods.
We note that \THISWORK\ is implemented on a CPU, and still has an average speedup of 28$\times$ over high-end NVIDIA GPU implementations. We do not give results for the motifs $M_{6-4}$ and $M_{6-5}$
since previous methods timeout even after running for a day. \THISWORK\ is able to get approximate
counts even for this motifs. On the specific money-laundering bipartite motif in \Fig{money-laundering}, \THISWORK\ takes \emph{4 minutes} on an example dataset, while the best exact method
takes \emph{two days}.

{\bf Low error of \THISWORK:} Despite the variety of complex constraints in \Fig{temp_motifs}, in \emph{all} cases, \THISWORK\ has significantly lower
error. We plot the errors
of previous work and \THISWORK\ in \Fig{error}.
\THISWORK\ typically has an error of less than 5\%, while previous methods
have 20\% error or more.
The only hard motifs for \THISWORK\ are $M_{5-4}$ and $M_{5-5}$, which have more than 20\% error. But other methods have twice as much (or much more) error.

{\bf Theoretical analysis of \THISWORK:} We give a comprehensive theoretical analysis of \THISWORK,
giving bounds on all running time steps, and an analysis of the sample complexity with respect
to output error.

\section{High-level Ideas In \THISWORK}
\label{sec:high-level}



The final \THISWORK\ algorithm is fairly complex with multiple moving parts. In this section,
we give an overview of various procedures in \THISWORK. We will denote the input graph as $G$
and the motif as $M$. A depiction of the \THISWORK\ pipeline is given in \Fig{overall-process}. 
There are two primary steps.

{\bf Preprocessing:} This is the main novelty in \THISWORK. We first choose an appropriate
spanning tree of $M$ and identify of a set of \emph{relaxed temporal constraints}. 
This process itself requires computations involving the input graph $G$. 
Our aim is to select a temporal "submotif" $T$ of $M$ that can be quickly sampled.
We set up the temporal constraints to allow for a dynamic programming bottom-up
approach to sample uniform random instances of $T$. This approach requires computing
a series of edge weights, called the \emph{sampling weights}.
Since there are multiple temporal edges between the same pair of vertices, 
the edge weights can be complicated to compute. 

{\bf Sampling:} These weights allow us to apply a common paradigm 
in approximate motif counting that involves sampling a substructure,
and extending to the motif~\cite{seshadhri2014wedge,jha2015path,wang2017moss,bressan2018motif,PaBh+24}.
We show how the weights can be used to subsample subtrees of $T$, which
can then be incrementally extended to a uniform random sample of $T$. 
A quick validation verifies that the sample
 satisfies the temporal constraints of $M$. The next step
is to count the number of matches to $M$ that involve the sampled match to $T$.
Again, temporal graphs bring in challenges. In the typical
setting, a single $T$ can only extend to a small number of matches to $M$. 
Here, the sampled $T$ could have many temporal edges going between the vertices,
and hence an unbounded number of matches to $M$ come from a \emph{single} sampled $T$.
We devise an algorithm using a careful binary searching over various edge lists
to compute this number of matches. 
The sampling step is repeated for a collection of samples. Standard randomized
algorithms techniques tell us how to rescale the total number of matches found,
to get an estimate for the total count of $M$.

We go deeper into the preprocessing step, which requires many new ideas.





\subsection{A Closer Look at Preprocessing} \label{sec:pre}



Consider mining the temporal motif $M_\text{5-3}$ as shown in \Fig{constraints}.
We fix the specific spanning tree shown in the figure.
The original motif imposes an ordering constraint on the edges and a total time window constraint.
It is not clear how to sample such a spanning tree, so we relax some of the constraints.

We relax the requirements for all edges to be ordered correctly and to fall within a specified time interval from \Def{temp-match}. These constraints
are only applied to certain pairs of \textit{adjacent edges}.
This forms the base constraint (Constraint 1), as illustrated in \Fig{constraint_1}.
In complex spanning trees, these constraints leave out multiple
ordering conditions.

We note that most efficient mechanisms to sample trees use dynamic programming,
and technically sample tree homomorphisms (where non-adjacent pattern vertices can be matched to 
the same vertex in $G$).
To address this, we introduce a straightforward constraint ensuring that the end vertices of adjacent edges are distinct. This approach minimizes memory overhead and significantly reduces invalid samples, making it particularly effective in skewed graphs. This is described as Constraint 2 is illustrated in \Fig{constraint_2}.

Constraints 1 and 2 focus on adjacent edges, while Constraint 3 introduces a global constraint of the time interval.
For example, in the chosen spanning tree of $M_\text{5-3}$, the sequential timing conditions of Constraint 1 implies $t_0 < t_4 \leq t_0 + 3\delta$, potentially violating the $\delta$ time interval constraint. \Revision{(We use $t_i$ to denote the timestamp the edge labeled $i$.) }
To reduce the affect of these violations, we apply a $2\delta$ sliding window approach, partitioning the input graph into overlapping subgraphs based on timestamps. This results in intervals like $[0, 2\delta]$, $[\delta, 3\delta]$, and so on, ensuring that any temporal motif matches within a $\delta$-time window must reside entirely within a single interval and will not span across the boundary of two intervals. This approach effectively reduces the time window length to $2\delta$, significantly reducing the occurrence of invalid matches due to $\delta$-window violations. An illustration of Constraint 3 is provided in \Fig{constraint_3}. Note that this approach leads to the under/over counting of some
temporal motifs due to boundary conditions. Since this overcounting factor can 
be determined for any match, we can apply corrections to the final estimate.

All in all, these constraints create a subset of matches for the spanning tree. We prove that
by a multi-pass preprocessing over the graph, we can construct an efficient sampler
for these matches of the spanning tree. This process involves a dynamic program that constructs
weights for various subtrees, and then computes weights for larger subtrees. The constraints
discussed above allow for this dynamic programmin approach to work.

{\em Choose a Spanning Tree:}
The first step of preprocessing requires selecting the best spanning tree. As shown in Figure~\ref{fig:spanning_trees}, a motif can have multiple valid spanning tree candidates.
We try to determine the spanning tree
that will lead to the fewest number of samples required for convergence.
\Revision{This is an important distinction from previous spanning tree based
sampling works that fix a choice~\cite{seshadhri2014wedge,jha2015path,wang2017moss, PaBh+24}. Those methods could afford to do that because they deal with smaller and
non-temporal motifs.}
We need to design some heuristics to select a smaller pool of spanning trees,
and then exactly determine the sampling cost for these trees.
By selecting the tree with the lowest sampling cost, we ensure efficient and accurate motif counting during execution. The details are elaborated in \Sec{heuristics}.

\section{Related Work}
\noindent \textbf{Static Motif Mining.}
There have been a lot of research in static motif mining, including exact counting and enumeration approaches \cite{chen2022efficient, jamshidi2020peregrine, jamshidi2021deeper,mawhirter2019graphzero, mawhirter2019automine, shi2020graphpi, teixeira2015arabesque, wei2022stmatch}, and approximate algorithms \cite{schank05finding, seshadhri2014wedge, jha2015path, liu2019sampling, sarpe2021presto, wang2020efficient, tsourakakis2009doulion, ahmed2014graph, turk2019revisiting, pavan2013counting, jha2015path, wang2017moss, jain2017fast, ye2022lightning, ye2023efficient, yang2021efficient, bressan2019motivo}
Although static motif mining can serve as a first step for temporal motif matching, as illustrated by Paranjape \textit{et al.}~\cite{paranjape2017motifs}, this often results in orders of magnitude redundant work ~\cite{mackey2018chronological, yuan2023everest} due to the presence of temporal constraints. 

\noindent \textbf{Exact Temporal Motif Mining.}
Temporal motif mining was first formally introduced by Paranjape \textit{et al.}~\cite{paranjape2017motifs}, who suggested an algorithm to list and count instances of temporal motifs. A newer exact counting method~\cite{mackey2018chronological} introduced a backtracking algorithm by listing all instances on edges sorted chronologically. Everest~\cite{yuan2023everest} further refines this approach, employing system-level optimizations on GPUs to enhance performance substantially. There are also numerous algorithms designed for general temporal motif counting~\cite{sun2019tm,min2023time} and others tailored to specific motifs~\cite{boekhout2019efficiently,kumar20182scent,pashanasangi2021faster,gao2022scalable}.
Despite these advancements, scalability remains a significant issue.

\noindent \textbf{Approximate Temporal Motif Mining.}
To approximate temporal motif counts, various sampling-based methods balance efficiency and accuracy.
There are several general frameworks for estimating temporal motifs.
\Revision{
Our work differs fundamentally from those methods such as IS~\cite{liu2019sampling}, PRESTO~\cite{sarpe2021presto}, and ES~\cite{wang2020efficient} in how it handles motif structure and temporal constraints.
IS~\cite{liu2019sampling} partitions the timeline into non-overlapping intervals and performs exact motif counting on a sampled subset of intervals. Its performance is highly sensitive to interval size and may miss cross-interval patterns.
PRESTO~\cite{sarpe2021presto} improves upon IS by avoiding rigid partitioning and instead using uniform sampling across the entire graph. While more flexible, it still treats motif instances in a uniform sampling space and does not incorporate structural information.
ES~\cite{wang2020efficient} samples individual edges uniformly and enumerates all local motif instances around them. It does not exploit the motif’s topology beyond local neighborhoods, leading to inefficiencies and high variance for larger motifs.
In contrast, TIMEST introduces a spanning tree–based sampling framework that leverages the motif’s structural backbone and performs weighted sampling along constrained subspaces.} 

There are methods that rely on an exact counting for some sampled graph, such as
Ahmed \textit{et al.}~\cite{ahmed2021online} and OdeN~\cite{sarpe2021oden}. 
These are often less efficient and have large estimator variance.
There are methods tailored to the unique characteristics of specific motifs.
Cai \textit{et al.}~\cite{cai2023efficient} and Pu \textit{et al.}~\cite{pu2023sampling} specialize in butterfly motifs in bipartite graphs, Oettershagen \textit{et al.}~\cite{oettershagen2023temporal} explore 2- and 3-node motifs and Pan \textit{et al.}~\cite{PaBh+24} focus on 4-node motifs.

\section{The Algorithmic Details of \THISWORK}
\label{sec:algorithm}



\begin{table}[t]
    \scriptsize
    \caption{Summary of notations}
    \vspace{-3mm}
    \label{tab:notation}
    \begin{tabular}{L{1.5cm}|L{6.2cm}}
         Symbol & Definition \\ \hline\hline
        $\delta$ & maximum time window \\ 
        $M=(H,\pi, \delta)$ & temporal motif \\
        $G$ & input graph with $m=|E(G)|$ edges and $n=|V(G)|$ vertices\\
        $\phi_V, \phi_E$ & vertex map $\phi_V: V(H) \to V(G)$ and edge map $\phi_E: E(H) \to E(G)$\\
        \Revision{$t(e)$} & \Revision{time stamp of edge $e$} \\
        $S$ & spanning tree of a temporal motif \\
        $P$ & sampled edges from $G$ that is mapped to $S$ \\
        $w_{s,e}$ & sampling weight of edge $e$ in $G$ that is mapped to $s$ in $S$\\
        $W$ & total sampling weights \\
        $C$, $\hat{C}$ & ground truth and estimated motif count \\
    \end{tabular}
\end{table}

In line with previous work~\cite{mackey2018chronological,yuan2023everest}, we store the graph $G$ as a sorted list of incoming and outgoing edges by timestamp for efficient binary search operations.
For ease of algorithm explanation, we adjust the timestamps of the edges in $G$ to begin at time 0.
We summarize the important symbols in \Tab{notation}.

We start with some important definitions for temporal graphs.

\begin{definition} [Temporal outlists and degrees]
\label{def:in_out_edges}
\Revision{For a vertex $v$ and timestamp $t$ and $t'$}, the \emph{temporal outlist} $\outlist{v}{t}{t'}$ is the set of outedges $(v,w,t^{\prime\prime})$, where $w$ is any vertex and $t^{\prime\prime} \in [t,t']$. 
The \emph{temporal inlist} $\inlist{v}{t}{t'}$ is defined similarly.

With $\alpha \in \{+, -\}, \beta \in \{<, >\}$ and fixed $\delta$, we use the notation $\alphalist{\alpha}{v}{t}{\beta}{\delta}$ to represent $\Lambda^{\alpha}_{v}[t-\delta, t]$ when $\beta$ is $<$, and $\Lambda^{\alpha}_{v}[t,t+\delta]$ when $\beta$ is $>$.

The \emph{temporal out-degree} $\tempout{v}{t}{t'}=|\outlist{v}{t}{t'}|$. And the \emph{temporal in-degree} $\tempin{v}{t}{t'}=|\inlist{v}{t}{t'}|$.
\end{definition} 

\begin{definition} [Multi-edge list]
\label{def:multiedge}
\Revision{We define multi-edge list $El_{u,v}[t,t']$ as the collection of edges $e=(u, v, t^{\prime\prime})$ with $t^{\prime\prime} \in [t, t']$.}

To simplify the notation with $\alpha \in \{+, -\}, \beta \in \{<, >\}$ and $\delta$, we use $\multiedgelist{\alpha}{{u,v}}{t}{\beta}{\delta}]$ to present different cases. Here, $\alpha$ represents the edge direction: $+$ means $u \to v$ direction, and $-$ means $v \to u$ direction. $\beta$ specifies the time range: the timestamp must fall in $[t-\delta, t]$ (<) or fall in $[t, t+\delta]$ (>).
\end{definition}

\begin{definition} [Multiplicity]
\label{def:multiplicity} 
Given a pair $u, v$ of distinct vertices and timestamps $t < t'$, the \emph{multiplicity} $\sigma_{u,v}[t,t']$ is the number of edges $(u,v,t^{\prime\prime})$ which satisfy $t^{\prime\prime} \in [t,t']$. $\sigma_{u,v}[t, t'] = |El_{u,v}[t,t']|$.
We denote the maximum $\delta$-multiplicity as $\sigma_\delta$, 
defined as $\max_{u,v,t}$ $\sigma_{u,v}[t,t+\delta]$. 
\end{definition}

Let $S$ be a spanning tree of the motif we wish to count. 
We first define the notion of a leaf edge. An edge $s \in S$ is called a leaf edge if any one of its endpoints has 0 indegree and outdegree. 

%



We introduce the process of assigning weights to edges $S$. 
It will be convenient to root the tree at an edge (rather than a vertex),
and we define parent/children using this rooting. Note that this rooting
is independent of the actual directions of the edges.
The height of a leaf edge is set to zero.  The height of an edge is the length
of the longest path down the tree to a leaf.
(Equivalently, the height of an edge is one plus the maximum height of its children.)

\begin{definition} \label{def:dep}
    The dependency list of an edge $s \in E(S)$ consists of triples  $\langle s', \alpha, \beta \rangle$, where $s' \neq s$ is a child of $s$, the parameter $\alpha$ (+ or -) determines the direction (incoming or outgoing) of $s'$ with respect to the vertex at which they meet and $\beta$ (<, >) gives the relative time-order between $s'$ and $s$. We denote this list by $D(s)$.
\end{definition}

Consider the tree $S$ in \Fig{overall-process}, where the edges $s_i$ is the edge
labeled $i$. Then, $D(s_1) = \{\langle s_0, +, < \rangle$\}.

\begin{definition} \label{def:delta-sp-tree}
    A spanning tree of a temporal motif $M$ is represented as a tuple $S = (L, O, D)$, where (i) $L$ is the list of edges in $E(H)$ that form the directed spanning tree of $M$; $(ii)$ $O$ is an ordering of the edges in $L$; $(iii)$ $D$ is the dependency of the edges in $L$. 
\end{definition}

The order $O$ can be any topological sorting based on the dependency $D$.
Now we introduce the notion of a \textit{partial match} to a tree in $G$, which will be central throughout the paper.
\begin{definition}
    A subgraph $H$ of $G$ is a $\delta$-partial match to tree $S$ 
    if there exists a pair $\phi = (\phi_V, \phi_E)$ of mappings such that  $\phi_V:V(S)\rightarrow V(G)$ and $\phi_{E}: E(S) \rightarrow E(G)$ with the following properties:
    \begin{enumerate}
        \item $\phi_{E}(u_i, u_j) = (\phi_V(u_i), \phi_V(u_j))$ 
        \item $(u_i, u_j) \in E(S)$ if and only if $(\phi_V(u_i), \phi_V(u_j)) \in E(G)$, with $\phi_V(u) \neq \phi_V(v)$
        \item Suppose $e_1 = (u,v)$ and $e_2 = (x,y)$ are dependent edges in $S$. 
        Assume that the height of $e_1$ exceeds the height of $e_2$. Then, if $t(u,v) < t(x,y)$, then $t(\phi_{E}(x,y)) \in [t(\phi_{E}(u,v)), t(\phi_{E}(u,v)) + \delta ]$. Otherwise, $t(\phi_{E}(x,y)) \in [t(\phi_{E}(u,v)) - \delta, t(\phi_{E}(u,v))]$.
        \item Further $|\phi_{E}(u,v) \cap \phi_{E}(x,y)| = 1$
    \end{enumerate}
\end{definition}
In particular, a partial match $\phi$ is a homomorphism (property $(2)$) of $S$ into $G$, which respects the relative time order between pairs of adjacent edges along every path from the root of $S$ to any of its leaves (property (3), corresponds to Constraint 1). It also ensures that matches to adjacent edges intersect only in one vertex (property (4), corresponds to Constraint 2).  Given a partial match $\phi$, an edge $e = (u,v,t) \in E(G)$, and an edge $s = (x,y) \in E(S)$, we say that $\phi$ matches $s$ to $e$ if $\phi_{E}(x,y) = (u,v)$. We will often refer to $e$ as the \emph{match} of $s$ in $G$ under $\phi$. 

\Revision{\textbf{Overview:} \Fig{overall-process} illustrates the overall process of estimating temporal motif counts for a given motif M in an input temporal graph G. The algorithm begins by selecting the best spanning tree of motif M to optimize sampling efficiency (\Sec{heuristics}), then determines relaxed temporal constraints (\Sec{pre}) and precomputes edge sampling weights (\Sec{preprocess}). In the sampling phase, it repeatedly samples spanning trees from G (\Sec{sample}), validates them against relaxed constraints, and derives local counts using dynamic programming (\Sec{valid-and-derive}). The final estimate is obtained by scaling the aggregated local counts from K samples (\Sec{overall-estimate}). We explain the spanning tree selection at the end,
since it requires many concepts from the other components.}

\subsection{Preprocess Sampling Weights}
\label{sec:preprocess}

During the preprocessing phase, for each edge of the input graph, we compute a set of $|E(S)|$ non-negative weights, one for every one of the $|E(S)|$ edges in the spanning tree $S$. 
We also refer to the root edge of $S$ as the center edge.

Furthermore, for an edge $s \in S$, $v_s$ denotes the lower-level endpoint of the edge $s$ and $T_s$ to denote the subtree rooted at $v_s$.

\begin{definition} [$s$-weight of edge $e$ $w_{s,e}$]
\label{def:weight}
    Given an edge $e = (u,v,t) \in E(G)$ and an edge $s \in E(S)$, the $s$-weight of $e$ is defined as the number of partial matches (in $G$) $\phi$ to the subtree $T_s$ which are rooted at $\phi_V(v_s)$, where $v_s$ the end-point of $s$ with lower height. 
    We will denote the $s$-weight of an edge $e$ by $w_{s,e}.$
\end{definition}

Consider $S$ in \Fig{overall-process}. Suppose  $e = (v_2, v_3, 40) \in E(G)$ is a match to $s_1 = (B, C)$. Then $\phi_v(B) = v_2$, $\phi_V(C) = v_3$. In this case, $w_{e,s_1}$ is the number of partial matches in $G$ to the subtree $T_B$ that are rooted at $v_2$. Since the subtree $T_B$ consists of a single edge $(A, B)$ that happens before $(B,C)$, $w_{e,s_1}$ assume a value equal to the number the inedges of $v_2$ with timestamps in the interval $[40 - \delta, 40]$.


More generally, suppose $e = (u,v,t) \in E(G)$ is a match to the edge $s = (x,y)$ under $\phi_E$. Consider an arbitrary triple $\langle s_1, \alpha_1, \beta_1 \rangle$ in the list $ D(s)$. Assume w.l.o.g. $\phi_V(v_s) = u$. We claim that any match to $s_1$ in $G$ under $\phi_E$ will either be an in-edge or out-edge of $u$ (depending on $\alpha_{1}$),  that does not intersect $v$, and whose timestamp is either in $[t - \delta, t]$ or $[t, t + \delta]$, depending on $\beta_{1}$. We denote this list by $L_{e,s,s_{1}}$, which is formally defined below.

\begin{claim}\label{clm:matching-edges-set}
    The partial match $\phi_E$ will map $s_1$ to an edge in  \\$\Lambda_{u}^{\alpha_1}[t, (\beta_1, \delta)]\setminus El_{uv}[t, (\beta_1, \delta)] $ if $\phi_V(v_s) = u$.

    The partial match $\phi_E$ will map $s_1$ to an edge in  \\$\Lambda_{v}^{\alpha_1}[t, (\beta_1, \delta)]\setminus El_{vu}[t, (\beta_1, \delta)] $ if $\phi_V(v_s) = v$.
\end{claim}

Thus, when looking for the potential matches to $s_1$, it suffices to restrict to the edge list given in \Clm{matching-edges-set}. Let us denote this list $L_{e,s,s_{1}}=\Lambda_{u}^{\alpha_1}[t, (\beta_1, \delta)]\setminus El_{uv}[t, (\beta_1, \delta)]$ (when $\phi_V(v_s) = u$) or $\Lambda_{v}^{\alpha_1}[t, (\beta_1, \delta)]\setminus El_{vu}[t, (\beta_1, \delta)]$ (when $\phi_V(v_s) = v$) for simplicity. 

Suppose the list $D(s)$ contains $k$ triples $\langle s_1, \alpha_1, \beta_1\rangle, \cdots, \langle s_k, \alpha_{k}, \beta_{k} \rangle$. Then any partial match $\phi$ to the subtree $T_{v_{s}}$ can be obtained by first picking a match to each of the $k$ edges $s_{1}, s_{2}, \dots, s_{k}$, and then extend them recursively into partial matches to the subtrees $T_{s_{1}}, \cdots T_{s_{k}}$.


\begin{claim}\label{clm:overall-num-matches}
    Suppose the list $D(s)$ contains  triples $<s_i, \alpha_i, \beta_i>$ for $1 \leq i \leq k$. Then \\
    $w_{s,e} = \Bigl(\displaystyle \sum_{e \in L_{e,s,s_{1}}}w_{s_{1},e}\Bigr) \Bigl(\displaystyle \sum_{e \in L_{e,s,s_{1}}}w_{s_{2},e}\Bigr)  \dots \Bigl(\displaystyle \sum_{e \in L_{e,s,s_{k}}}w_{s_{k},e}\Bigr)$ 
\end{claim}
\begin{proof}
    Let Match($e = (u,v,t),s$) denote the collection of partial matches to the subtree $T_s$, which are rooted at $\phi_V(v_{s})$ in $G$. Here $\phi_v(v_s) \in \{u,v\}$. Then
    Match($e,s$) = $\{(\phi_1, \dots, \phi_k)\}$, where $\phi_i$ is a partial match to the subtree $T_{s_i}$ in $G$. Every such match  is rooted at a suitable end-point of some edge $e_i \in L_{e,s,s_i}$. This endpoint is  given by $\phi_{Vi}(v_{s_i})$. 

    Note that the $s$ weight of $e$ is precisely the size of the set Match($e,s$) i.e. $w_{s,e} = |\text{Match}$$(e,s)|$. 

    Now recall that the number of partial matches to the subtree $T_{s_{i}}$, which can be obtained from an edge $e$ in the list 
$L_{e,s,s_i}$ is exactly equal to $e$'s $s_i$ weight. Hence the number of partial matches $\phi_i$ with the given property is $\displaystyle \sum_{e \in L_{e,s,s_i}}w_{s_i,e}$. 

As a result, we can conclude that the number of tuples in the set Match($e,s$) is given by \\ $\Bigl(\displaystyle \sum_{e \in L_{e,s,s_{1}}}w_{s_{1},e}\Bigr) \Bigl(\displaystyle \sum_{e \in L_{e,s,s_{1}}}w_{s_{2},e}\Bigr)  \dots \Bigl(\displaystyle \sum_{e \in L_{e,s,s_{k}}}w_{s_{k},e}\Bigr)$. This completes the proof.
\end{proof}

\Clm{overall-num-matches} allows us to use dynamic programming to efficiently compute the $w_{s,e}$ values, where the choices for $s$ vary in the increasing order of their height in $S$.

Finally, let $W$ denote the total number of partial-matches to $S$ in $G$ and $c$ be the center edge of $S$. Then :

\begin{claim}\label{clm:num-matches-entire-tree}
 $W = \sum_{e \in E(G)}w_{c,e}$, where $c$ is the center edge in $S$
\end{claim}
\begin{proof}
     Since any edge in $G$ could potentially be the center edge in a match to $S$, it follows that the total number of matches to $S$ in $G$ is $\sum_{e \in E(G)}w_{c,e}$. 
This completes the proofs.
\end{proof}
    


\begin{algorithm}[t]
\caption{: \textsc{Preprocess}(S)}
\label{algo:preprocess}
\begin{flushleft}
\textbf{Input}: Spanning tree $S=(L, O, D)$  \\
\textbf{Output}: Total sampling weight $W$; $W_i$ for each subgraph $G_i$; $w_{i, s, e}$ for $e$ in $G_i$ that are mapped to $s$. 
\begin{algorithmic}[1]
\State{$num_\text{subg}$ = time span of $G$ / $\delta$} 
\For{$i \in [0, num_\text{subg})$} \Comment{partition to subgraphs} 
\State{build subgraph $G_\text{i}$ whose timestamp in $[i\delta, (i+2)\delta]$}
\State{$W_{i}, w_{i, s, e}$ = \textsc{PreprocessSubgraph}($S, G_{i}$)}
\EndFor
\State \Return $W, W_{i}, w_{i, s, e}$
\end{algorithmic}
\end{flushleft}
\end{algorithm}

\begin{algorithm}[t]
\caption{: \textsc{PreprocessSubgraph}($S, G_\text{i}$)}
\label{algo:preprocessSubgraph}
\begin{flushleft}
\textbf{Input}: Spanning tree $S=(L, O, D)$, and current subgraph $G_\text{i}$  \\
\textbf{Output}: $W$ for the current subgraph $G_\text{i}$; $w_{s, e}$ for edge $e\in G_\text{i}$ mapped to edge $s \in L$.
\begin{algorithmic}[1]
\For{$h \in [1, H]$} \Comment{skip leaf edges in $S$}
\For{$e = (u,v,t) \in E(G_\text{i})$}
\For{$s \in O[h]$}
\State calculate $w_{s,e}$ according to \Clm{overall-num-matches}
\If{s is the "center" edge in $S$}
\State $W \pluseq w_{s, e}$
\EndIf
\EndFor
\EndFor
\EndFor
\State \Return $W, w_{s, e}$
\end{algorithmic}
\end{flushleft}
\end{algorithm}

We detail the \textsc{Preprocess} procedure in \Alg{preprocess}. First, we segment the entire timespan of the input graph $G$ into overlapping intervals of length $2\delta$. Let $I = \{[0, 2\delta], [\delta, 3\delta], \dots, [(q-2)\delta, q\delta] \}$ denote the resulting collection of intervals. Here $q = T/\delta$ where $T$ is the time span of $G$. 
Each interval $I_i$ corresponds to a subgraph $G_i$, consisting of the edges in $G$ which fall within the interval $I_i$.
Next, we invoke the subroutine \textsc{PreprocessSubgraph} to get the overall sampling weight $W_{i}$ for each subgraph $G_i$, as well as the $s$-weights for the edges in each subgraph. Here $s$ varies across the edges of $S$.

In the \textsc{PreprocessSubgraph} procedure (in \Alg{preprocessSubgraph}), we compute the $s$-weight for every edge $e \in E(G)$ and every $s \in E(S)$. These weights are stored in a 2D array and are computed in the order given by $O$. 
More concretely, for every $e \in E(G)$, first we compute its $s$-weight for every $s \in E(S)$ which is at \emph{height 1} in $S$. (Leaf edges with 0 height are skipped because the weights are constant 1s.) 
Next, for each edge, we compute its associated $s$-weight for every $s \in E(S)$ which is at \emph{height 2}. We use the relation given by \Clm{overall-num-matches} to do so.  
We repeat these steps until we reach the center edge in $S$. 
The total sampling weight for the current subgraph is then calculated by \Clm{num-matches-entire-tree}.

\begin{algorithm}[t]
\caption{: \textsc{SampleSubg}($S, G_{i}, W, w_{s, e}$)}
\label{algo:sampleSubg}
\begin{flushleft}
\textbf{Input}: Spanning tree $S=(L, O, D)$, sampling weights $W$ for current subgraph $G_{i}$, and $w_{s,e}$ for edges in $E(G_{i})$ \\
\textbf{Output}: Sampled edges $P$ where $P[i]$ is the edge in $E(G_{i})$ that is mapped to $s_i$ in $S$, partial match vertex map $\phi_V$ and edge map $\phi_E$.
\begin{algorithmic}[1]
\State {$c$ is the center edge in $S$}
\State Sample center edge $e$ with sampling probability $p_{e} = w_{c,e} / W_{i}$\label{line:sample-center-edge}
\State $P[c] = e$, Update $\phi_V$ and $\phi_E$
\For{$h \in [H, 0]$}
\For{$s \in O[H]$}
\For{$\langle s', \alpha', \beta' \rangle \in D(s)$}
\State Use binary search to find the $L_{e,s,s'}$ \label{line:sample-candidate-list}
\State $W_x=\sum_{e \in L_{e,s,s'}}{w_{s', e}}$
\State Compute $p_{s',e} = w_{s', e} / W_x$ for $e \in L_{e,s,s'}$ \label{line:sample-distribution}
\State Sample $e_{s'}$ using sampling weight $p_{s',e}$
\State $P[s'] = e_{s'}$, update $\phi_V$ and $\phi_E$
\EndFor
\EndFor
\EndFor
\State \Return $P$, $\phi_V$ and $\phi_E$
\end{algorithmic}
\end{flushleft}
\end{algorithm}
\subsection{Sample}
\label{sec:sample}
The goal of the \textsc{Sample} procedure is to sample a partial match to spanning tree $S$ from graph $G$. First, we sample an interval from $I$ with probability proportional to its overall weight $W_i$. Next, we construct the graph $G_i$ by including those edges of $G$ whose timestamp belongs to the sampled interval $I_i$. 

Then we invoke the \textsc{SampleSubg} function on $G_i$ (\Alg{sampleSubg}). 
In this subroutine, first we sample a match $e = (u,v,t) \in E(G_i)$ to the center edge $c$ of $S$. This is done by sampling an edge $e \in E(G)$ with probability proportional to $w_{c,e}$ (line~\ref{line:sample-center-edge}). Next, for every $s \in D(c)$, we use binary search to get the list of potential matches to $s$ in $G$. For each $s \in D(c)$, we sample its match in $G$ by picking an edge from the list of candidate matches to $s$ with probability proportional their associated $s$-weight. Having done this for every $s \in D(c)$, we recursively build a match to $S$, now by sampling matches to the edges in $D(s)$ for every $s \in D(c)$. Note that the order in which we sample a match to the spanning tree is exactly the opposite of how we compute the edge weights.
Let $\mathcal{G} := \{ G_1, G_2, \dots, G_{|I|}\}$  be the family of subgraphs of $G$ obtained from the intervals in $I$. For a fixed $G_k \in \mathcal{G}$ containing $\phi$, our algorithm outputs a uniform random partial match to $S$. Refer to \Lem{sample-subtree-overall} for details.
\begin{lemma}\label{lem:sample-subtree-overall}
    Fix a partial match $\phi$ to $S$ in $G$.  If this partial match is present in $N_\phi$-many subgraphs of $\mathcal{G}$, then the subroutine \textsc{SampleSubg} will output $\phi$ with probability $N_\phi/W$. 
\end{lemma}
Note that $N_\phi \leq 2$ for every partial match $\phi$ to $S$. For a fixed $G_k \in \mathcal{G}$ containing $\phi$, our algorithm outputs a uniform random partial match to $S$.

\subsection{Validate and Derive the Motif Counts}
\label{sec:valid-and-derive}
\begin{algorithm}[t]
\caption{: \textsc{ValidateAndDeriveCnt}($M, P, \phi_E, \phi_V$)}
\label{algo:validateAndDeriveCnt}
\begin{flushleft}
\textbf{Input}: Temporal motif $M$, sampled spanning tree $P$, partial vertex map $\phi_V$ and edge map $\phi_E$ \\
\textbf{Output}: Number of motif instances $cnt$ that extend from $\phi_E(S)$
\begin{algorithmic}[1]
\State Verify if $\phi_V$ is a 1-1 map. If not, \Return 0
\State Verify if edges in $\phi_E(S)$ are in $\delta$ time range. If not, \Return 0
\State Verify if edges in $\phi_E(S)$ have the correct edge order. If not, \Return 0
\State \Return \textsc{DeriveCnt}$(M, P, \phi_E, \phi_V) / N_{\phi}$
\end{algorithmic}
\end{flushleft}
\end{algorithm}

\begin{algorithm}[t]
\caption{: \textsc{DeriveCnt}$(M, P, \phi_E, \phi_V)$)}
\label{algo:DeriveCnt}
\begin{flushleft}
\textbf{Input}: \textcolor{black}{Temporal motif $M$, sampled spanning tree $P$, partial vertex map $\phi_V$ and edge map $\phi_E$} \\
\textbf{Output}: \textcolor{black}{Number of motif instances $cnt$ that extend from $\phi_E(S)$}
\begin{algorithmic}[1]
\State \textcolor{black}{For every edge $e$ of $M$ that's not in the sampled spanning tree $P$, use binary search to find the lists (in G) of potential matches}
\State \textcolor{black}{Number these lists as $L_1, L_2, \cdots L_l$ in the time ordering.}
\State  \textcolor{black}{Use the \textsc{ListCount} algorithms (Alg. 4 of~\cite{pan2024accurate}) to count all possible combinations without enumeration.}
\State \Return \textcolor{black} {cnts}
\end{algorithmic}
\end{flushleft}
\end{algorithm}

After sampling a partial match $\phi$ to $S$, we verify if it respects all ordering and temporal constraints. 
If the sampled match fails to meet any required conditions, we set its associated count to 0. Otherwise, we obtain the number of instances of the target motif induced by it. This can be done by a combination of the merge-technique used in the merge-sort algorithm and dynamic programming \textsc{DeriveCnt} (\Alg{DeriveCnt}), which uses the \textsc{ListCount} algorithm (Alg. 4 from Pan \textit{et al.}~\cite{pan2024accurate}). \Revision{Given a collection of time ordered edge lists $L_1, L_2, \cdots L_l$, it counts the number
of combinations $(e_1, e_2, \ldots, e_l)$ where for all $i$, $e_i \in L_i$, 
and $t(e_i) < t(e_{i+1})$. This is done with pointer traversals and dynamic programming. Note that by modifying the \textsc{DeriveCnt} function to list the sampled motif instance to reconstruct individual motif instances, our algorithm can be adapted to support random motif listings. Finally, we rescale this count to ensure that our estimate is unbiased at the end. The full procedure is detailed in \Alg{validateAndDeriveCnt}.}

\begin{algorithm}[t]
\caption{\textsc{Estimate}$(M,G,k)$}
\label{algo:overall}
\begin{flushleft}
\textbf{Input}: Temporal motif $M$, input graph $G$, and sample number $k$ \\
\textbf{Output}: Motif count estimate $\widehat{C}$
\end{flushleft}
\begin{algorithmic}[1]
\State Use heuristic to choose a spanning tree $S$
\State $W, W_{i}, w_{i, s, e}$ = \textsc{Preprocess}$(S)$
\State{$num_\text{subg}$ = time span of $G$ / $\delta$} 
\State{init sample counts $samp_{i} = 0$ for every subgraph $G_{i}$, $cnt = 0$}
\For{$j \in [1, k]$} \Comment{sample subgraphs}
\State{Sample $G_{i}$ with probability $p_{i, \delta} = W_{i} / W$}
\State{$samp_{i} \pluseq 1$}
\EndFor
\For{$i \in [1, num_\text{subg}]$}
\For{$j \in [1, samp_{i}]$}  \Comment{sample edge in subgraphs}
\State{$P, \phi_E, \phi_V$ = \textsc{SampleSubg} ($S, G_{i}, W_{i}, w_{i, s, e}$)}
\State $cnt \pluseq \textsc{ValidateAndDeriveCnt}(M, P, \phi_E, \phi_V)$ \label{algo:overall-checkMotif}
\EndFor
\EndFor
\State \Return $\widehat{C} = (cnt/k)\cdot W$
\end{algorithmic}
\end{algorithm}

\subsection{Overall Estimate Procedure}
\label{sec:overall-estimate}
Integrating the procedures described previously, we outline the  \textsc{Estimate} process in \Alg{overall}.
First, analyze the motif and use heuristics to choose a spanning tree $S$ of the motif. Given $S$, for every $s \in E(S)$ and $e \in E(G)$, we compute the weight $w_{s,e}$, as given by the \textsc{Preprocess} subroutine. Next, we invoke the \textsc{SampleSubg} procedure to sample a partial match to $S$ and feed it as input to the \textsc{ValidateAndDeriveCnt} subroutine, which in turn returns the number of matches to the target motif induced by the partial match to $S$. We run the procedure multiple times, average the individual counts, rescale the average and return the final result.


Next, we show that the estimate produced by our algorithm is unbiased and bounded. 

\begin{lemma}\label{lem:unbiased-estimate}
     Let $\widehat{C}$ denote the estimate produced by TIMEST and $C$ be the true motif count. Then $\EX[\widehat{C}] = C$.
\end{lemma}
\begin{proof}
    Let $\mathcal{S}$ denote the set of partial matches to the spanning tree $S$ in $G$. For every $1 \leq i \leq k$, let $\phi_i$ denote the partial match to $S$ output by the subroutine \textsc{SampleSubg} when invoked for the $i$-th time. For every $\phi \in \mathcal{S}$, let $M_{\phi}$ be the number of instances of the target motif $H$ that extend from $\phi$, $N_\phi$ be the number of subgraphs in $\mathcal{G}$ which contain $\phi$  and $C$ be the total number of occurrences of $H$ in $G$. Further, let the random variable $Y_i$ denote the  outcome of \textsc{Derivecnt}  when invoked on $\phi_i$ and let $Y = \sum_{i \leq k}Y_i$. 

    $\EX[Y] = \EX[\sum_{i \leq k}Y_i] = \sum_{i \leq k} \EX[Y_i]$. \\
    Note that $\EX[Y_i] = \sum_{\phi \in \mathcal{S}}\EX[Y_i|\phi_i =\phi]\Pr(\phi_i = \phi)$. 
    
    From \Lem{sample-subtree-overall}, $\Pr(\phi_i = \phi) = N_\phi/W$. Further, $\EX[Y_i|\phi_i = \phi] = M_{\phi}/N_{\phi}$. This is because once we obtain a partial match $\phi$, we determine the number of subgraphs in $\mathcal{G}$ that contain $\phi$ i.e. $N_{\phi}$. Subsequently, we divide the value returned by the \textsc{Derivecnt} subroutine when invoked on $\phi$ with $N_\phi$.
    
    Therefore,  $\EX[Y_i] = \sum_{\phi \in \mathcal{S}}(M_\phi/N_\phi)(N_\phi/W) = \sum_{\phi \in \mathcal{S}}(M_\phi/W) = C/W$. Note that $\sum_{\phi \in \mathcal{S}}M_{\phi} = C$ because every instance $H_G$ of $H$ in $G$ has a unique spanning tree $T_{H_G}$ such that the subroutine \textsc{DeriveCnt} will account for $H_G$ only when it is invoked with $T_{H_G}$ as its input.

    Therefore, $\EX[Y] = kM/W$. So, $\widehat{C} = WY/k$,  is an unbiased estimate of $C$.
\end{proof}
We state the following theorem, which will help us in proving concentration results regarding our estimate.

\begin{theorem} \label{thm:chernoff} [Theorem 1.1 of~\cite{DuPa-book}] Let $Y_1, Y_2, \ldots, Y_k$ be independent random variables in $[0,1]$, and let $Y = \sum_{i \leq k} Y_i$. Let $\varepsilon \in (0,1)$. Then, 
$$ \Pr[|Y - \EX[Y]| > \varepsilon \EX[Y]] \leq 2 \exp(-\varepsilon^2 \EX[Y]/3)$$
\end{theorem}
We now prove the following result.
Let $B$ denote the maximum number of matches to the target motif $M$ resulting from matches to $S$ in $G$. 

\begin{theorem}\label{thm:convergence-estimate}
   Let  $Y_1, Y_2, \dots, Y_k$ be $k$ random variables, where $Y_i$ is the outcome of the subroutine \textsc{ValidateAndDeriveCnt} on the $i$-th sampled partial match to $S$. Suppose $k = (3B/\varepsilon^{2})(W/C)\ln(2/\gamma)$ for any $\varepsilon, \gamma \in (0,1)$. Then with probability at least $1 - \gamma$, $\widehat{C} \in [(1 - \varepsilon)C, (1 + \varepsilon)C]$. 
\end{theorem}
\begin{proof}
Recall that the random variables $Y_i$ denote the outcome of the \textsc{Derivecnt}  when invoked on partial match $\phi_i$ and  $Y = \sum_{i \leq k}Y_i$. 
   First, note that the $k$ random variables $Y_1, \dots, Y_k$ are independent of each other. Also, observe that the random variables $Y_1, \dots, Y_k$ may not necessarily assume values in the interval $[0,1]$. Hence we cannot directly use the Chernoff Bound. However, dividing them by $B$ ensures that $0 \leq Y_i/B \leq 1$  for all $1 \leq i \leq k$.

Before applying the Chernoff Bound, observe that the event $|Y/B - kM/BW| \geq (\varepsilon kM/WB)$ is identical to $|WY/k - C| \geq \varepsilon C$. Similarly, since $\widehat{C} = WY/K$, the events $|Y/B - kM/WB| \geq (\varepsilon kM/WB )$ and $|\widehat{C} - C| \geq \varepsilon C$ are also identical.

Note that $\EX[Y/B] = \sum_{i \leq k}\EX[Y_i]/B = Mk/WB$. Using Chernoff Bound,  for any $\varepsilon \in (0,1)$, $\Pr(|Y/B - (kM/WB)| \geq (\varepsilon kM/WB)) \leq 2\exp(-(\varepsilon^{2}/3)(kM/WB))$. Since 
$k = (3B/\varepsilon^{2})(W/C)\ln(2/\gamma)$, it follows that $\Pr(|Y/B - (kM/BW)| \geq (\varepsilon kM/BW)) \leq \\ 2\exp(-(\varepsilon^{2}/3)(3B/\varepsilon^{2})(W/C)(\ln(2/\gamma))(C/WB)) = \gamma$.
\end{proof}
We present the time complexity of procedure \textsc{Preprocess} and \textsc{SampleSubg} as follows.
\begin{claim} \label{clm:time-complexity}
        The \textsc{Preprocess} procedure takes $O(|E(S)|^2 m d_{\max})$ time. The \textsc{SampleSubg} procedure takes $O(|E(S)|^{2}d_{\max})$ time per sample. \Revision{The storage complexity is $O(m|V(S)|)$}. Here, $m$ is the number of edges in $G$, where $d_{\max}$ is the maximum number of simple edges incident on any vertex in $G$.
\end{claim}

\begin{proof}
    The \textsc{PreprocessSubgGraph} function computes the $s$-weights ($w_{s,e}$) for every $e \in E(G)$ and every $s \in E(S)$. In order to determine the value of $w_{s,e}$, we first do  binary searches on edges incident on a suitable endpoint of $e$ (say $u$) to determine potential matches to the edges in $S$ which contribute to the list $D(s)$. Each such search takes $O(\log m)$ time. Subseqeuently, for every $s' \in D(s)$, we add the $s'$-weights of the matches to $s'$ incident on $u$ and take their product to get the $s$-weight of $e$. Since $s'$ weights for each edge would already by pre-computed and stored, we spend $O(1)$ time to fetch those weights for the edges matching every $s' \in D(s)$. So it takes $O(|E(s)|d_{\max} + \log m)$ time to compute $w_{s,e}$ value. Since we repeat this procedure for every $e \in E(G)$ and $s \in E(S)$, we get the overall time complexity to be $O(|E(S)|^{2}md_{\max})$.  

    During the sampling phase, once we have sampled an edge $e$ that matches $s$, we use binary search to determine potential matches to every $s' \in D(s)$, fetch their corresponding $s'$-weights and subsequently sample matches to every $s'$ from their corresponding list of potential matches using distributions defined using $s'$-weight values. So the first step takes $O(|E(S)|\log m)$ time, the second step takes $O(|E(S)|d_{\max})$ and final step takes $O(|E(S)| \log m)$ time. We do these steps for every $s$ in $E(S)$, so the overall time to sample an entire partial match is $O(|E(S)|^{2}d_{\max})$ per sample. 
    

    \Revision{Regarding the space complexity, we store the sampling weight $w_{s,e}$ in a vector for each edge in the graph $G$, whose size equal to the total edge count ($m$) multiplying the number of edges in the spanning tree of the motif ($|V(S)|$).}
\end{proof}

\subsection{Choosing the Spanning Tree}
\label{sec:heuristics}

\begin{figure}
    \centering
    \includegraphics[width=0.8\linewidth]{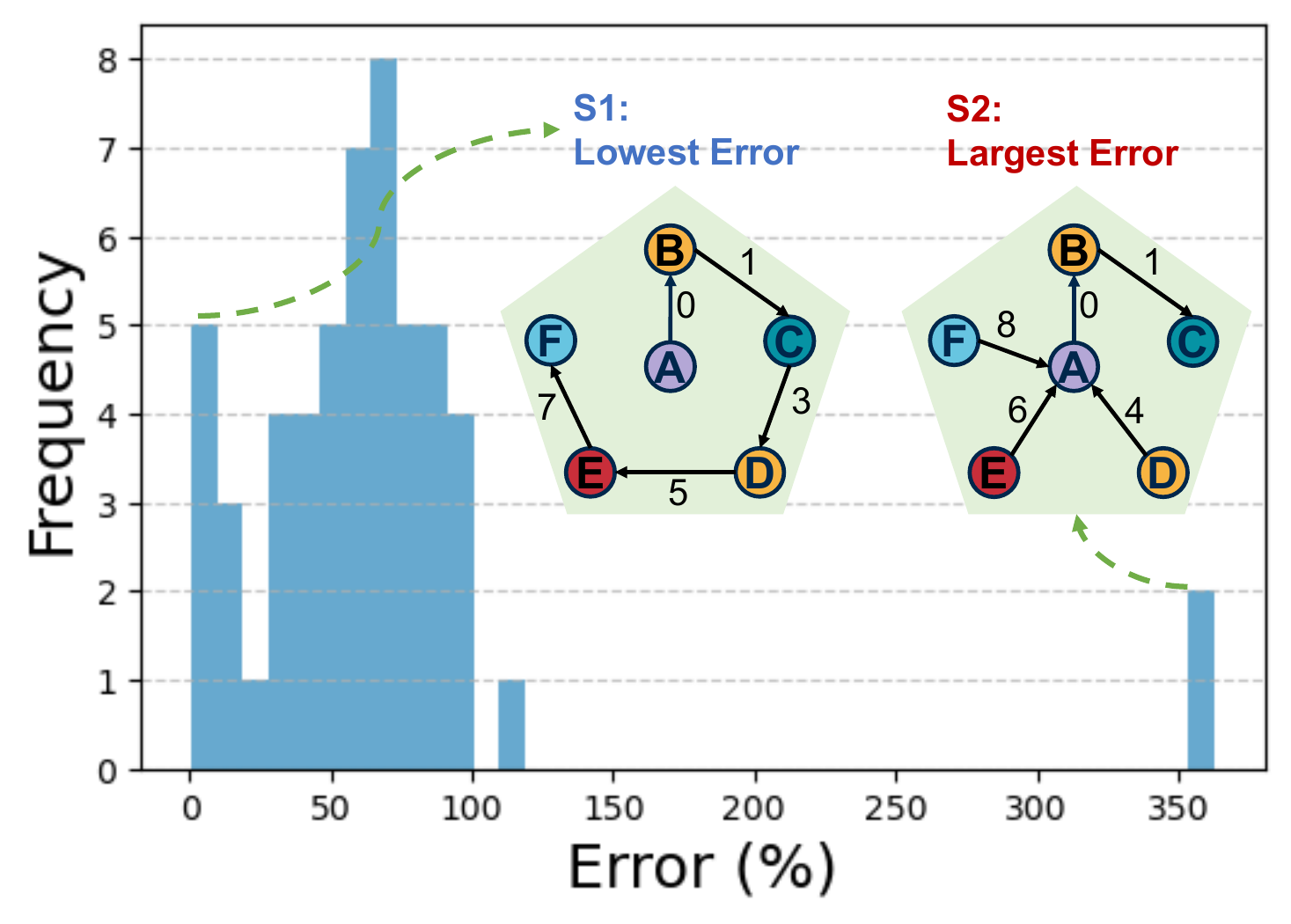}
    \vspace{-4mm}
    \caption{Histogram of estimation errors (\%) across different spanning tree choices for motif $M_{6-4}$ and with 1e8 samples. Two highlighted examples, S1 and S2, correspond to the lowest and highest observed errors, emphasizing that the choice of spanning tree significantly impacts estimation accuracy.}
    \label{fig:hist_err_sptrees}
\end{figure}

\begin{algorithm}[t]
\caption{: \textsc{getBestSPTree}($M, G, n_c$)}
\label{algo:heuristics}
\begin{flushleft}
\textbf{Input}: Temporal motif $M$, input graph $G$, number of candidate spanning trees $n_c$ \\
\textbf{Output}: Selected spanning tree $S_{best}$
\begin{algorithmic}[1]
\State DFS to enumerate all spanning trees of $M$ in an array $S_{all}[]$
\State $C\_losseness = \textsc{getConstraintLooseness}(S_{all}, M)$
\State $S_{topk}$ = \textsc{selectTopK}($C\_losseness, S_{all}, n_c$)
\State Init an array $est\_runtime[n_c]$
\For{$i \in [0, n_c)$}
\State $curW$ = \textsc{PreprocessSubgraph}$(S_{topk}[i], G)$
\State $est\_runtime[i] = curW$
\EndFor
\State $S_{best}$ = tree with min $est\_runtime$ \\
\Return $S_{best}$
\end{algorithmic}
\end{flushleft}
\end{algorithm}

\begin{algorithm}[t]
\caption{: \textsc{getConstraintLooseness}($S_{all}, M$)}
\label{algo:getSPTreeStats}
\begin{flushleft}
\textbf{Input}: all spanning trees $S_{all}[]$ \\
\textbf{Output}: Constraint looseness $C\_lossness[]$
\begin{algorithmic}[1]
\State Initialize $C\_lossness[]$ with 0
\For{$i \in [0, len(S_{all}))$}
\State \texttt{/*For every node, compute the absdiff of the edges incident to the node*/}
\For{$u \in nodes(M)$}
\State $E(u) \gets$ edges if $M$ that are incident on $u$
\If{$|E(u)| < 2$}
\State continue
\EndIf
\State \texttt{/*Consider all pairs of incident edges*/} 
\For{each unordered pair $(e_1, e_2)$ in $E(u)$}
\State $C\_lossness[i] += |t(e_1) - t(e_2) - 1|$
\EndFor
\EndFor
\EndFor
\\
\Return $C\_lossness$
\end{algorithmic}
\end{flushleft}
\end{algorithm}


\Fig{spanning_trees} shows that a motif can have many spanning trees. 
Our goal is to select the spanning tree that minimizes runtime while having the best estimation quality.
The estimation error when choosing 54 different spanning trees of the $M_{6-4}$ motif is shown in \Fig{hist_err_sptrees}. The results reveal large variation in accuracy: the worst-case estimation error exceeds 350\% while the best error is less than 1\%. 





In \Thm{convergence-estimate}, given the motif $M$ and input graph $G$, the number of samples $k$ to take to reach convergence is proportional to the total sampling weight $W$. This weight $W$ 
is exactly the number of instances of chosen spanning tree $S$ in the graph. A smaller $W$ indicates a rarer spanning tree, reducing the search space. For example, consider the candidate spanning trees $S_1$ and $S_2$ of motif $M_{5-3}$ in \Fig{spanning_trees}, or those of motif $M_{6-4}$ in \Fig{hist_err_sptrees}. $S_1$ follows a path-like structure  and $S_2$ has a more tree-like structure. The experiments in \Tab{SPTree} show that the sampling weight $W$ for $S_2$ is around 5-10 times larger than that for $S_1$. This indicates that $S_1$ appears less frequently in the input graph and introduces stricter constraints compared to $S_2$, so we prefer $S_1$ to $S_2$.

However, calculating $W$ for \textit{all} spanning trees is time-consuming. Each calculation takes approximately 10\% of the overall runtime.  We propose a heuristic that approximates how tightly a spanning tree enforces temporal constraints, which serves as a rough proxy for $W$ without accessing the graph. This proxy, denoted as $C\_looseness$, is computed in \Alg{getSPTreeStats}. It examines all edge pairs incident to each node and computes how tightly the spanning tree orders them. A lower value indicates stricter constraints and is thus preferred.


Overall, our heuristics have the following steps.
\begin{asparaitem}
\item Finding Candidates: We use DFS to find all spanning trees and evaluate them based on motif-only information. 
We select the top $n_c$ candidates with the smallest constraint loss (\textsc{getConstraintLooseness}).
\item Evaluating Candidates: For the selected candidate spanning trees, we calculate the actual sampling weight using the input graph (\textsc{PreprocessSubgraph}) . A smaller W means fewer samples are required for convergence (indicated by \Thm{convergence-estimate}).
\item Choosing the Best Tree: We use the sampling weight W (which affects the number of samples) to estimate the total runtime. We select the spanning tree with the smallest estimated runtime, making it applicable across different graphs and motifs. 
\end{asparaitem}

\section{Evaluation}

In this section, we comprehensively evaluate the performance of \THISWORK\ across different graphs, motifs, and $\delta$ values. We assess both the accuracy and the runtime of \THISWORK and compare them with existing methods, highlighting its accuracy and efficiency.
\subsection{Experiment Setup}

\begin{table}[t]
\small
\caption{Temporal graph datasets used in the evaluation.}
\vspace{-3mm}
\label{tab:dataset}
\centering
\scriptsize
\begin{tabular}[htbp]
{P{1.8cm}|P{0.8cm}|P{1.4cm}|P{1cm}}
\hline
\textbf{Dataset} & \textbf{$|V|$} & \textbf{$|E_{temporal}|$} & \textbf{Time span (year)} \\ \hline \hline
wiki-talk (WT) & 1.1M & 7.8M & 6.4\\ \hline
stackoverflow (SO) & 2.6M & 63.5M &  7.6 \\ \hline
bitcoin (BI) & 48.1M & 113.1M & 7.1\\ \hline
reddit-reply (RE) & 8.4M & 636.3M & 10.1 \\\hline
\end{tabular}
\end{table}

\noindent\textbf{Benchmarks.}
We evaluate the temporal motif counts across a spectrum of datasets, encompassing medium to large-scale graphs such as wiki-talk (WT), stackoverflow (SO)~\cite{snapnets}, bitcoin (BI)~\cite{Kondor-2014-bitcoin}, and reddit-reply (RE)~\cite{hessel2016science}, detailed in \Tab{dataset}. 
To showcase the versatility of \THISWORK, we mine temporal motifs encompassing 5-vertex and 6-vertex connected graphs in \Fig{temp_motifs}. We set $\delta$ as $8W$, $16W$ for WT and SO, and as $1D$ for BI and RE, where $H$ represents hour, $D$ represents day, and $W$ represents week. 

\noindent\textbf{Baselines.}
For exact algorithms, we use the state-of-the-art Everest~\cite{yuan2023everest}, implemented in\cite{Everest_impl}, which builds on the BT algorithm~\cite{mackey2018chronological} with GPU-specific optimizations. We exclude the original BT CPU version due to its significantly lower performance (over 20× slower than Everest). 

\Revision{For approximate algorithms on 4-vertex motifs, we compare against PRESTO~\cite{sarpe2021presto}, ES~\cite{wang2020efficient} and IS~\cite{liu2019sampling}.} PRESTO is a sampling framework that uniformly samples intervals of length $c\delta$ and performs exact temporal motif counting on these sampled intervals to derive estimated counts. We explore two versions of PRESTO, referred to as \textit{PRESTO-A} and \textit{PRESTO-E}. 
\Revision{For 5- for 6-vertex motifs, we omit the comparison with IS~\cite{liu2019sampling} as PRESTO was already shown to outperform it in their study. And we do not provide a quantitative comparison with  ES~\cite{wang2020efficient} because the public code~\cite{ES_impl} is limited to motifs with 4 or fewer edges}.

\noindent\textbf{\THISWORK}
The parameter that affects the accuracy and runtime of \THISWORK\ is the number of samples $k$, which varies from 10 million to 1 trillion. For the detailed settings per case, please refer to the \code{reproduce.py} in \url{https://anonymous.4open.science/r/TIMEST/}.
\Revision{For a fair comparison, the runtime of \THISWORK\ include all steps in \Fig{overall-process}, which consists of preprocess and sample procedure.}

\noindent\textbf{Accuracy Metrics}
The accuracy of approximate algorithms is defined as $\frac{|C-\hat{C}|}{C}$, where $C$ and $\hat{C}$ are the exact count and the estimated count, respectively. And error = 1- accuracy. We run it 5 times to get the average(avg) and standard deviation(std) of error.

\noindent\textbf{Discussion Of The Samples}
\THISWORK\ uses more samples than prior work, but the runtime per sample
is extremely small (i.e., 0.1 microsecond per sample). PRESTO samples
time windows and then runs an expensive algorithm on this sample. Each sample in \THISWORK\ is a spanning tree that processes few edges. So, even including
preprocessing time, \THISWORK\ is faster.


\noindent\textbf{Hardware Platforms.}
Both \THISWORK\ and PRESTO run on a CPU platform with 32 threads. Specifically, the CPU platform utilized is an AMD EPYC 7742 64-core CPU with 64MB L2 cache, 256MB L3 cache, and 1.5TB DRAM memory. Everest~\cite{Everest_impl} runs on a single NVIDIA A40 GPU with 10k CUDA cores and 48GB memory.


\subsection{Results}

\noindent\textbf{\THISWORK\ is fast.}
In \Tab{runtime-BT-this}, we list the runtime of Everest (GPU) and \THISWORK\ (CPU) on counting the 5-vertex and 6-vertex temporal motifs in \Fig{temp_motifs}. We focus on the cases where Everest runs more than 60 seconds. 
For RE dataset with $\delta$=1 day, Everest is faster than \THISWORK.
With large number of nodes and edges in input graph and motifs, Everest takes a large amount of runtime, and even timeout (>1 day of execution time) in some cases.
On the other hand, \THISWORK\ completes execution in less than 30 minutes in all cases and exhibits 
a geomean speedup of 28$\times$ over Everest.

As a further demonstration,
we also \THISWORK\ and Everest (GPU) on the bipartite money laundering pattern of \Fig{money-laundering}~\cite{altman2024realistic}. 
We use the WT graph with $\delta$ set to 4W. Everest takes two days to complete, while \THISWORK\
runs in \emph{four minutes}. The error observed is less than $0.6\%$.

\begin{table}[ht]
\scriptsize
\centering
\caption{Runtime (in second) of Everest (GPU) and \THISWORK\ (CPU), the speedup of \THISWORK\ over Everest, and the estimation error (\%) of \THISWORK. We set the timeout limit as 1 day. We mark the speedup and error as \textit{No Exact} for the case that Everest timeout. The error(\%) is represented as avg $\pm$ std.}
\label{tab:runtime-BT-this}
\begin{tabular}{c|c|c|P{1cm}|P{1.2cm}|P{1cm}|c}
\textbf{Graph} & \textbf{$\delta$} & \textbf{Motif} & \textbf{Everest (GPU)} & \textbf{\THISWORK\ (CPU)} & \textbf{Speedup ($\times$)} & \textbf{Error(\%)} \\ \hline \hline
\multirow{10}{*}{WT} &\multirow{10}{*}{8W} &$M_\text{5-1}$ &9.9E1 &9.3E0 &10.6 &0.1$\pm$0.0 \\ 
& &$M_\text{5-2}$ &1.7E3 &8.8E0 &193.3 & 1.3$\pm$0.7 \\ 
& &$M_\text{5-3}$ &1.5E3 &7.1E1 &21.1 & 0.7$\pm$0.5 \\ 
& &$M_\text{5-4}$ &2.5E3 &7.1E1 &34.9 & 6.4$\pm$4.9 \\
& &$M_\text{5-5}$ &3.4E3 &4.3E2 &8.0 & 9.4$\pm$5.3 \\
& &$M_\text{6-1}$ &1.8E4 &1.0E2 &176.1 & 0.0$\pm$0.0 \\ 
& &$M_\text{6-2}$ &1.2E4 &1.4E1 &852.9 & 0.1$\pm$0.1 \\
& &$M_\text{6-3}$ &4.1E3 &3.3E2 &11.8 & 0.7$\pm$0.6 \\
& &$M_\text{6-4}$ &5.5E4 &1.3E3 &42.5 & 7.4$\pm$3.8 \\
& &$M_\text{6-5}$ &6.8E4 &1.3E3 &51.8 & 72.3$\pm$16.7 \\ \hline
\multirow{10}{*}{SO} &\multirow{10}{*}{8W} &$M_\text{5-1}$ &9.0E2 &3.0E1 &30.5 &0.0$\pm$0.0 \\
& &$M_\text{5-2}$ &7.7E3 &2.9E1 &265.0 & 0.9$\pm$0.3\\
& &$M_\text{5-3}$ &6.6E2 &8.0E1 &8.2 & 0.8$\pm$0.4 \\
& &$M_\text{5-4}$ &4.0E2 &8.1E1 &5.0 & 20.4$\pm$12.6 \\
& &$M_\text{5-5}$ &6.7E3 &6.4E2 &10.5 & 32.7$\pm$16.7 \\
& &$M_\text{6-1}$ &3.1E4 &4.2E1 &734.8 & 0.1$\pm$0.1 \\
& &$M_\text{6-2}$ &9.4E4 &3.2E1 &2980.1 & 0.2$\pm$0.0 \\
& &$M_\text{6-3}$ &2.1E3 &1.5E2 &13.6 & 0.3$\pm$0.2 \\
& &$M_\text{6-4}$ &2.9E3 &1.0E3 &2.8 & 33.1$\pm$18.5 \\
& &$M_\text{6-5}$ &1.5E5 &1.0E3 &140.1 &100.0$\pm$0 \\ \hline
\multirow{10}{*}{BI} &\multirow{10}{*}{1D} &$M_\text{5-1}$ &3.7E2 &5.4E1 &6.8 &0.0$\pm$0.0 \\
& &$M_\text{5-2}$ &2.1E3 &4.9E1 &43.1 &2.2$\pm$0.7 \\
& &$M_\text{5-3}$ &1.9E3 &1.1E2 &16.7 &1.4$\pm$0.6 \\
& &$M_\text{5-4}$ &7.7E3 &1.1E2 &68.3 &3.2$\pm$3.1 \\
& &$M_\text{5-5}$ &1.0E4 &8.9E2 &11.3 &10.0$\pm$5.4 \\
& &$M_\text{6-1}$ &5.1E4 &6.2E1 &812.7 &0.1$\pm$0.0 \\
& &$M_\text{6-2}$ &timeout &4.7E1 &No Exact & No Exact \\
& &$M_\text{6-3}$ &3.0E3 &1.4E2 &20.9 &0.2$\pm$0.2 \\
& &$M_\text{6-4}$ &timeout &8.3E2 &No Exact & No Exact \\
& &$M_\text{6-5}$ &timeout &8.2E2 &No Exact & No Exact \\ \hline
\multirow{10}{*}{RE} &\multirow{10}{*}{1D} &$M_\text{5-1}$ &6.3E1 &3.9E2 &0.2 &0$\pm$0 \\
& &$M_\text{5-2}$ &3.4E2 &3.9E2 &0.9 &0.6$\pm$0.4 \\
& &$M_\text{5-3}$ &5.5E2 &4.1E2 &1.4 &0.4$\pm$0.5 \\
& &$M_\text{5-4}$ &9.6E3 &4.2E2 &23.1 &2.0$\pm$1.3 \\
& &$M_\text{5-5}$ &9.9E3 &1.3E3 &7.8 &19.9$\pm$10.6 \\
& &$M_\text{6-1}$ &5.1E2 &5.1E2 &1.0 &0.1$\pm$0.0 \\
& &$M_\text{6-2}$ &3.1E3 &3.9E2 &7.9 &0.0$\pm$0.0 \\
& &$M_\text{6-3}$ &1.0E3 &8.5E2 &2.2 &0.1$\pm$0.0 \\
& &$M_\text{6-4}$ &timeout &8.5E2 &No Exact & No Exact \\
& &$M_\text{6-5}$ &timeout &8.7E2 &No Exact & No Exact \\
\bottomrule
\end{tabular}
\end{table}

\begin{table}[t]
\centering
\scriptsize
\caption{Runtime (in second) and relative error (\%) of approximate algorithms on various datasets and temporal motifs. The lowest error is highlighted in \colorbox{Green1}{green} block. All run in 32-thread. We set the timeout limit as (1 day). If the program runs out of memory (OOM), it will be killed by Linux. \THISWORK\ is the fastest and the most accurate in most cases.}
\label{tab:approx}
\begin{tabular}{P{0.8cm}|P{0.6cm}|P{0.8cm}|P{0.6cm}|P{0.8cm}|P{0.6cm}|P{0.8cm}|P{0.6cm}}
\hline
\multirow{2}{*}{ \textbf{Dataset} } & \multirow{2}{*}{ \textbf{Motif} } & \multicolumn{2}{c|}{\textbf{PRESTO-A}} & \multicolumn{2}{c|}{\textbf{PRESTO-E}} & \multicolumn{2}{c}{\textbf{\THISWORK}} \\ 
\cline{3-8}
& & \textbf{Time (s)} & \textbf{Error} & \textbf{Time (s)} & \textbf{Error} & \textbf{Time (s)} & \textbf{Error} \\ \hline\hline
\multirow{7}{*}{\shortstack{WT\\ $\delta=8W$}} & $M_\text{5-1}$ & 8.0E3 & 3.0\% & 4.0E4 & 7.1\% & 1.0E1 & \cellcolor{Green1}\textbf{0.1\%} \\ \cline{2-8}
& $M_\text{5-2}$ & 1.5E4 & 28.1\% & 1.6E4 & 2.2\% & 8.8E0 & \cellcolor{Green1}\textbf{1.3\%} \\ \cline{2-8}
& $M_\text{5-3}$ & 1.5E4 & 11.5\% & 1.9E4 & 12.0\% & 6.7E1 & \cellcolor{Green1}\textbf{0.7\%} \\ \cline{2-8}
& $M_\text{5-4}$ & 1.1E4 & 24.1\% & 5.1E4 & 33.6\% & 6.4E1 & \cellcolor{Green1}\textbf{6.4\%} \\ \cline{2-8}
 & $M_\text{5-5}$ & 2.4E4 & 25.8\% & 2.4E4 & 77.1\% & 4.8E2 & \cellcolor{Green1}\textbf{9.4\%} \\ \cline{2-8}
 & $M_\text{6-1}$ & 1.1E4 & 93.8\% & 2.0E3 & 98.8\% & 6.4E1 & \cellcolor{Green1}\textbf{0.0\%} \\ \cline{2-8}
& $M_\text{6-2}$ & 2.5E3 & 96.1\% & timeout & N/A & 1.1E1 & \cellcolor{Green1}\textbf{0.1\%} \\ \cline{2-8}
 & $M_\text{6-3}$ & 1.4E5 & 107.5\% & 2.6E5 & 10.4\% & 2.0E2 & \cellcolor{Green1}\textbf{0.7\%} \\ \cline{2-8}
 & $M_\text{6-4}$ & 1.3E4 & 89.5\% & timeout & N/A & 6.9E2 & \cellcolor{Green1}\textbf{7.4\%} \\ \cline{2-8}
 & $M_\text{6-5}$ & timeout & N/A & timeout & N/A & 6.9E2 & 72.3\% \\ \hline
\multirow{7}{*}{\shortstack{SO\\ $\delta=8W$}} & $M_\text{5-1}$ & 4.9E3 & 3.8\% & 2.3E4 & 17.2\% & 2.6E1 & \cellcolor{Green1}\textbf{0.0\%} \\ \cline{2-8}
 & $M_\text{5-2}$ & OOM & N/A & OOM & N/A & 2.5E1 & \cellcolor{Green1}\textbf{0.9\%} \\ \cline{2-8}
 & $M_\text{5-3}$ & 8.0E2 & 33.4\% & 1.0E3 & 17.2\% & 6.6E1 & \cellcolor{Green1}\textbf{0.8\%} \\ \cline{2-8}
 & $M_\text{5-4}$ & 4.0E2 & 87.3\% & 2.8E2 & 35.7\% & 6.9E1 & \cellcolor{Green1}\textbf{20.4\%} \\ \cline{2-8}
 & $M_\text{5-5}$ & 1.5E4 & 81.5\% & OOM & N/A & 4.8E2 & \cellcolor{Green1}\textbf{32.7\%}  \\  \cline{2-8}
 & $M_\text{6-1}$ & 8.6E4 & 98.4\% & timeout & N/A & 3.2E1 & \cellcolor{Green1}\textbf{0.1\%} \\ \cline{2-8}
 & $M_\text{6-2}$ & OOM & N/A & OOM & N/A & 2.9E1 & \cellcolor{Green1}\textbf{0.2\%} \\ \cline{2-8}
 & $M_\text{6-3}$ & 1.4E4 & 45.0\% & 4.0E4 & 13.88\% & 1.2E2 & \cellcolor{Green1}\textbf{0.3\%} \\ \cline{2-8}
 & $M_\text{6-4}$ & 1.3E3 & 81.4\% & 3.9E3 & 270.8\% & 8.7E2 & \cellcolor{Green1}\textbf{33.1\%} \\ \cline{2-8}
 & $M_\text{6-5}$ & OOM & N/A & OOM & N/A & 8.8E2 & 100\% \\ \hline
\multirow{7}{*}{\shortstack{BI\\ $\delta=1D$}} & $M_\text{5-1}$ & 4.1E2 & 12.9\% & 2.7E3 & 13.2\% & 3.9E1 & \cellcolor{Green1}\textbf{0.0\%} \\ \cline{2-8}
& $M_\text{5-2}$ & 8.2E2 & 47.3\% & 1.3E3 & 41.7\% & 3.6E1 & \cellcolor{Green1}\textbf{2.2\%}\\ \cline{2-8}
& $M_\text{5-3}$ & 1.2E2 & 65.9\% & 3.8E2 & 39.4\% & 4.1E1 & \cellcolor{Green1}\textbf{1.4\%}\\ \cline{2-8}
 & $M_\text{5-4}$ & 7.7E2 & 41.1\% & 5.4E2 & 53.3\% & 9.6E1 & \cellcolor{Green1}\textbf{3.2\%} \\ \cline{2-8}
 & $M_\text{5-5}$ & 1.4E3 & 15.3\% & 2.8E3 & 11.5\% & 7.8E2 & \cellcolor{Green1}\textbf{10.0\%} \\ \cline{2-8}
 & $M_\text{6-1}$ & 5.5E4 & 52.4\% & timeout & N/A & 5.2E1 & \cellcolor{Green1}\textbf{0.1\%} \\ \cline{2-8}
 & $M_\text{6-2}$ & timeout & N/A & timeout & N/A & 5.2E1 & NE \\ \cline{2-8}
 & $M_\text{6-3}$ & 8.5E2 & 26.5\% & 5.9E3 & 32.2\% & 1.2E2 & \cellcolor{Green1}\textbf{0.2\%} \\ \cline{2-8}
 & $M_\text{6-4}$ & NE & NE & NE & NE & 7.9E2 & NE \\ \cline{2-8}
 & $M_\text{6-5}$ & NE & NE & NE & NE & 8.1E2 & NE \\ \hline
\multirow{7}{*}{\shortstack{RE\\ $\delta=1D$}} & $M_\text{5-1}$ & 3.2E2 & 102.5\% & 1.6E3 & 14.4\% & 3.2E2 & \cellcolor{Green1}\textbf{0.0\%} \\ \cline{2-8}
& $M_\text{5-2}$ & 3.7E2 & 92.8\% & 2.4E3 & 45.8\% & 2.9E2 & \cellcolor{Green1}\textbf{0.6\%}\\ \cline{2-8}
& $M_\text{5-3}$ & 1.8E4 & 925.0\% & 1.9E3 & 95.5\% & 4.2E2 & \cellcolor{Green1}\textbf{0.4\%}\\ \cline{2-8}
 & $M_\text{5-4}$ & 6.8E4 & 90.6\% & 5.6E4 & 97.6\% & 3.3E2 & \cellcolor{Green1}\textbf{2.0\%} \\ \cline{2-8}
 & $M_\text{5-5}$ & 1.1E4 & 98.6\% & 2.5E4 & 25.8\% & 1.2E3 & \cellcolor{Green1}\textbf{19.9\%} \\ \cline{2-8}
 & $M_\text{6-1}$ & 1.1E4 & 98.6\% & 4.5E4 & 99.9\% & 1.2E3 & \cellcolor{Green1}\textbf{0.1\%} \\ \cline{2-8}
 & $M_\text{6-2}$ & 3.6E4 & 40.4\% & 1.3E5 & 6.0\% & 3.5E2 & \cellcolor{Green1}\textbf{0.0\%} \\ \cline{2-8}
 & $M_\text{6-3}$ & 3.4E3 & 76.5\% & 1.1E3 & timeout & 4.5E2 & \cellcolor{Green1}\textbf{0.1\%} \\ \cline{2-8}
 & $M_\text{6-4}$ & NE & NE & NE & NE & 7.9E2 & NE \\ \cline{2-8}
 & $M_\text{6-5}$ & NE & NE & NE & NE & 8.1E2 & NE \\ \hline
\end{tabular}
\end{table}

\noindent\textbf{\THISWORK\ is accurate.}
Our comparison between \THISWORK\ and PRESTO~\cite{sarpe2021presto} focuses on  runtime performance and relative error, with findings summarized in \Tab{approx}. Despite PRESTO's runtime being 6$\times$ slower compared to \THISWORK, our results show that \THISWORK\ consistently outperforms PRESTO in accuracy for all evaluated temporal motifs involving 5 and 6 vertices.

PRESTO employs an approximation strategy that involves uniform sampling across graph partitions within a time window of size $c\delta$ ($c > 1$ is a parameter), and applying an exact algorithm on these partitions. 
PRESTO's inefficiency is because of the reliance on uniform sampling and the use of
the slow backtracking (BT) algorithm~\cite{mackey2018chronological} as the subroutine on the sampled time window. 

Both PRESTO and \THISWORK\ struggle to accurately estimate counts for the most complex 6-clique motif ($M_\text{6-5}$), showing low convergence and accuracy. Despite generating approximately 1 billion samples, with around 100 million being valid (i.e., not violating the constraints defined in \Def{temp-match}), only 1 - 100 of these valid samples contribute to the final count.
The 6-clique 
encompasses more constraints (such as edge mapping and ordering) than any chosen spanning tree can cover. 
Developing accurate algorithms for such complex motifs would require new ideas.

\begin{table*}[htbp]
\centering
\scriptsize
\caption{\Revision{Runtime (s) and relative error (\%) of all algorithms on each dataset. The lowest error is highlighted in \colorbox{Green1}{green} block. Because not all sampling algorithms support multi-threading, for a fair comparison, we report the runtime for single-threaded implementations. Because \THISWORK\ output count for $M_\text{4-1}$-$M_\text{4-4}$ at once, we consider the runtime of \THISWORK\ as the total runtime for $M_\text{4-1}$-$M_\text{4-4}$. \THISWORK\ is always the fastest.}}
\label{tab:4-vertex-runtime}
\begin{tabular}{P{1cm}|P{0.6cm}|P{0.9cm}|P{0.6cm}|P{0.9cm}|P{0.6cm}|P{0.9cm}|P{0.6cm}|P{0.9cm}|P{0.6cm}|P{0.9cm}|P{0.9cm}|P{0.6cm}}
\hline
\multirow{2}{*}{ \textbf{Dataset} } & \multirow{2}{*}{ \textbf{Motif} } & \multicolumn{2}{c|}{\textbf{PRESTO-A}} & \multicolumn{2}{c|}{\textbf{PRESTO-E}} & \multicolumn{2}{c|}{\textbf{IS}} & 
\multicolumn{2}{c|}{\textbf{ES}} & \multicolumn{3}{c}{\textbf{\THISWORK}} \\ \cline{3-13}
& & \textbf{Time (s)} & \textbf{Error} & \textbf{Time (s)} & \textbf{Error} & \textbf{Time (s)} & \textbf{Error} & \textbf{Time (s)} & \textbf{Error} & \textbf{Time (s)} & \textbf{Speedup over PRESTO-A ($\times$)} & \textbf{Error} \\ \hline \hline

\multirow{4}{*}{WT, $\delta$=8W} 
& M4-1 & 3.65E+03 & 31.08\% & 3.62E+03 & 8.65\% & 1.32E+04 & 1.62\% & 2.22E+02& 3.64\% & \multirow{4}{*}{ 5.16E+01 } & \multirow{4}{*}{ 292 } & \cellcolor{Green1}\textbf{0.26\%} \\
& M4-2 & 3.76E+03 & 4.69\% & 4.04E+03 & 7.92\% & 1.44E+04 & 0.80\% & 2.71E+02& 2.42\% & & & \cellcolor{Green1}\textbf{0.63\%} \\
& M4-3 & 3.65E+03 & 75.88\% & 3.96E+03 & 34.45\% & 1.32E+04 & \cellcolor{Green1}\textbf{0.57\%} & 2.35E+02& 8.34\% & & & 2.35\% \\
& M4-4 & 3.99E+03 & 77.56\% & 3.74E+03 & 27.54\% & 1.45E+04 & 12.63\% & 5.01E+02 & \cellcolor{Green1}\textbf{2.27\%} & & & 4.30\% \\ \hline

\multirow{4}{*}{ST, $\delta$=8W} 
& M4-1 & 3.69E+03 & 13.16\% & 4.10E+03 & 19.17\% & 7.99E+04 & 0.68\% & 1.87E+03 & 0.75\% & \multirow{4}{*}{ 3.70E+02 } & \multirow{4}{*}{ 41 } & \cellcolor{Green1}\textbf{0.24\%} \\
& M4-2 & 3.88E+03 & 86.36\% & 3.91E+03 & 12.84\% & 
 > 1D & - & 1.50E+03 & 0.50\% & & & \cellcolor{Green1}\textbf{0.37\%} \\
& M4-3 & 3.62E+03 & 38.68\% & 4.01E+03 & 20.29\% & 8.59E+04 & 1.45\% & 2.24E+02 & \cellcolor{Green1}\textbf{0.95\%} & & & 1.30\%\\
& M4-4 & 3.85E+03 & \cellcolor{Green1}\textbf{0.52\%} & 4.07E+03 & 10.05\% & 8.12E+04 & 0.56\% & 3.12E+02 & 1.76\% & & & 4.74\% \\ \hline


\multirow{4}{*}{BI, $\delta$=1D} 
& M4-1 & 3.62E+03 & 5.33\% & 3.60E+03 & 19.10\% & 1.80E+03 & 1.81\% & 8.66E+03 & 2.12\% & \multirow{4}{*}{ 4.60E+02 } & \multirow{4}{*}{ 32 } & \cellcolor{Green1}\textbf{0.07\%} \\
& M4-2 & 3.65E+03 & 18.50\% & 3.61E+03 & 8.94\% & 2.01E+03 & \cellcolor{Green1}\textbf{0.23\%} & 1.22E+03 & 1.48\% & & & 0.53\% \\
& M4-3 & 3.64E+03 & 25.39\% & 3.67E+03 & 19.33\% & 1.79E+03 & 20.56\% & 5.24E+02 & 6.23\% & & & \cellcolor{Green1}\textbf{1.84\%} \\
& M4-4 & 3.67E+03 & 41.71\% & 3.63E+03 & 31.99\% & 1.69E+03 & 16.61\% & 6.17E+02 & \cellcolor{Green1}\textbf{6.32\%} & & & 7.60\% \\ \hline


\multirow{4}{*}{RE, $\delta$=1D} 
& M4-2 & 3.60E+03 & 26.83\% & 3.61E+03 & 23.17\% & 3.67E+03 & 38.34\% & 1.85E+03 & 8.90\% & \multirow{4}{*}{ 2.42E+03 } & \multirow{4}{*}{ 6 } & \cellcolor{Green1}\textbf{0.22}\% \\
& M4-3 & 3.60E+03 & 160.88\% & 3.60E+03 & 53.66\% & 5.91E+03 & 21.37\% & 1.08E+03 & 1.45\% & & & \cellcolor{Green1}\textbf{0.44}\% \\
& M4-5 & 3.61E+03 & 89.03\% & 3.61E+03 & 46.35\% & 4.22E+03 & 33.11\% & 6.04E+02 & 2.96\% & & & \cellcolor{Green1}\textbf{1.92}\% \\
& M4-7 & 3.60E+03 & 104.42\% & 4.54E+03 & 344.62\% & 7.29E+03 & 151.53\% & 1.01E+03 & 27.94\% & & & \cellcolor{Green1}\textbf{4.57}\% \\ \hline

\hline
\end{tabular}
\end{table*}

\Revision{For smaller motifs with 4-vertex in \Fig{temp_motifs}, we compare \THISWORK\ with PRESTO, ES and IS baselines regarding runtime and eraltive error. The result is shown in \Tab{4-vertex-runtime}. As expected, \THISWORK\ is
always faster by an order of magnitude with comparable or lower error.}

\begin{table}[htb]
    \caption{The percentage of invalid and valid sample rate, relative error, and runtime when enforcing different constraints (C) for estimating temporal motif count of $M_\text{5-5}$.}
    \vspace{-3mm}
    \label{tab:constraints}
    \centering
    \scriptsize
    \begin{tabular}{c|c|c|c|c|c|c|c}
        \hline
\multicolumn{2}{c|}{Graph, $\delta$} & \multicolumn{3}{c|}{WI, $\delta$=8W} & \multicolumn{3}{c}{RE, $\delta$=1D} \\ \hline
\multicolumn{2}{c|}{Constraints}                  & C1      & C1+2    & C1+2+3   & C1       & C1+2     & C1+2+3    \\ \hline
\multirow{3}{*}{\shortstack{Invalid sample \\ rate (\%)}}   & vertex map & 42.9\%  & 36.3\%  & 36.6\%  & 75.5\%  & 2.0\%   & 1.9\%   \\ \cline{2-8}
& $\delta$ interval & 14.6\%  & 17.1\%  & 10.3\%  & 0.8\%   & 23.1\%  & 13.6\%  \\ \cline{2-8}
& edge order & 34.4\%  & 35.8\%  & 39.8\%  & 23.0\%  & 45.4\%  & 51.1\%  \\ \hline
\multicolumn{2}{c|}{\shortstack{Valid samples rate (\%)}} & 8.1\%   & 10.8\%  & 13.3\%  & 0.8\%    & 29.5\%   & 33.4\%   \\ \hline
\multicolumn{2}{c|}{Error (\%)}        & 5.1\%   & 0.9\%   & 2.9\%   & 95.92\%  & 12.5\%  & 4.0\%   \\ \hline
\multicolumn{2}{c|}{Runtime (s)}        & 320.1   & 332.7   & 361.0   & 1259.2  & 1221.7  & 1148.1  \\ \hline
    \end{tabular}
\end{table}

\noindent\textbf{Improved performance with additional constraints.}
Introducing specific constraints on sampled edges helps reduce the chances of them violating the requirements of \Def{temp-match}. 
In \Tab{constraints}, we show how the three constraints in \Sec{high-level} reduce invalid samples and improve performance. We categorize three distinct invalid sample types because of violating: a) 1-1 vertex map $\phi_V$ b) $\delta$ time interval c) edge orders. 

Our experimental approach employs Constraint 1 as the foundational condition, with Constraints 2 and 3 subsequently integrated in a phased manner. These experiments were conducted on the $M_\text{5-5}$ motif (5-cliques), utilizing the WT graph with $\delta = 8W$, and the RE graph with $\delta = 1D$. 
The objective is to achieve high accuracy with high efficiency, striving for an increase in the rate of valid samples while minimizing runtime for an equivalent number of samples.

From the table, we can see that adding Constraint 2 (distinct end vertices on dependent edges) reduces the number of invalid samples due to violating 1-1 vertex map. For RE graph, it significantly reduces the percentage of invalid samples due to violating the vertex map from 75\% to 2\%.
Adding Constraint 3 ($2\delta$ window) reduces the invalid samples due to violating the $\delta$ constraints, increasing the valid sample rate by around 5\%. 
Integrating all three constraints, we observe that the runtime remains largely unchanged, suggesting that the additional computational overhead introduced by these new constraints is minimal. More crucially, this integration boosts the valid sample rate from 1\% to 33\% for the RE graph, concurrently decreasing the error rate dramatically from 95\% to 4\%. \textit{This demonstrates not only the feasibility of applying these constraints without significant efficiency losses but also their effectiveness in enhancing the precision of the sampling process.}

\begin{table}[ht]
    \centering
    \scriptsize
    \caption{Sampling weight $W$, estimation error (\%) and runtime (s) for choosing different spanning tree ($S_1$ to $S_3$) when estimating motif count for $M_\text{5-3}$.}
    \label{tab:SPTree}
    \begin{tabular}{P{0.3cm}|c|P{0.45cm}|P{0.45cm}|P{0.45cm}|P{0.45cm}|P{0.45cm}|P{0.45cm}|c|c|c}
 \multirow{2}{*}{ \textbf{$G$} } & \multirow{2}{*}{ \textbf{$\delta$} }&\multicolumn{3}{c|}{\textbf{W}} &\multicolumn{3}{c|}{\textbf{Error}} &\multicolumn{3}{c}{\textbf{Runtime (s)}} \\ \cline{3-11}
 & &$S_1$ &$S_2$ &$S_3$ &$S_1$ &$S_2$ &$S_3$ &$S_1$ &$S_2$ &$S_3$ \\ \hline\hline
\multirow{2}{*}{WT} &4W &1.1E12 &8.9E12 &1.4E12 &\textbf{4.9\%} &9.5\% &14.5\% &\textbf{9.9} &12.4 &11.7 \\ \cline{2-11}
&8W &6.5E12 &5.2E13 &8.0E12 &\textbf{0.8\%} &13.4\% &3.4\% &\textbf{14.2} &16.4 &17.7 \\\hline
\multirow{2}{*}{SO} &4W &3.1E12 &2.6E13 &4.0E12 &\textbf{3.1\%} &33.1\% &5.8\% &25.4 &\textbf{19.8} &27.0 \\ \cline{2-11}
&8W &1.6E13 &1.5E14 &2.0E13 &\textbf{4.4\%} &21.8\% &11.4\% &30.0 &\textbf{20.7} &31.1 \\ \hline
BI &1D &2.7E13 &1.3E15 &1.2E14 &\textbf{1.2\%} &110.6\% &3.9\% &42.9 &\textbf{33.6} &48.1 \\ \hline
RE &1D &7.3E11 &4.9E12 &5.3E12 &\textbf{0.3\%} &3.5\% &6.8\% &327.3 &\textbf{251.5} &336.0 \\ \hline
\end{tabular}
\end{table}
\noindent\textbf{Effect of different spanning trees on performance.}
We examined using three distinct spanning trees for motif $M_\text{5-3}$, using 20 million samples each for our analysis, as seen in \Tab{SPTree}.
The trees $S_1$ and $S_2$ are displayed in \Fig{spanning_trees}. 

The total sampling weight $W$ is similar for $S_1$ and $S_3$ in certain graphs, but $S_2$ has a $W$ value about 10 times higher. 
\Thm{convergence-estimate} states that for a given graph and motif, the number of samples $k$ needed to reach convergence is proportional to $W$. Smaller $W$ implies the same error with fewer samples, which is corroborated
by \Tab{SPTree}.


Runtime comparison, given equal sample numbers, reveals minor differences between the trees, with $S_3$ having slightly longer runtimes than $S_1$ and $S_2$. Because the \textsc{ValidateAndDeriveCnt()} function has time complexity linear to the overall number of potential candidates (see \Sec{valid-and-derive}). Here the candidates refer to the non-spanning-tree edges of the motif. 



\begin{table}[htb]
    \centering
    \scriptsize
    \caption{Peak memory usage (GB) for PRESTO and \THISWORK.}
    \label{tab:peak_memory}
    \begin{tabular}{c|c|c|c|c}
         \textbf{Dataset} & \textbf{Motif} & \textbf{PRESTO-A} & \textbf{PRESTO-E} & \textbf{\THISWORK}\\ \hline \hline
         \multirow{3}{*}{\shortstack{WT\\ $\delta=8W$}} & $M_\text{5-1}$ & 1.5 & 3.7 & 2.9 \\ \cline{2-5}
         & $M_\text{5-3}$ & 15.9 & 37.2 & 2.8 \\ \cline{2-5}
         & $M_\text{6-3}$ & 2.7 & 6.9 & 3.0\\ \hline
         \multirow{3}{*}{\shortstack{SO\\ $\delta=8W$}} & $M_\text{5-1}$ & 12.4 & 13.1 & 20.1 \\ \cline{2-5}
         & $M_\text{5-3}$ & 98.1 & 111.6 & 20.1 \\ \cline{2-5}
         & $M_\text{6-3}$ & 26.8 & 30.8 & 21.0 \\ \hline
         \multirow{3}{*}{\shortstack{BI\\ $\delta=1D$}} & $M_\text{5-1}$ & 4.2 & 5.1 & 40.3 \\ \cline{2-5}
         & $M_\text{5-3}$ & 8.3 & 19.2 & 40.3 \\ \cline{2-5}
         & $M_\text{6-3}$ & 5.7 & 11.9 & 42.1 \\ \hline
    \end{tabular}
\end{table}
\noindent\textbf{Peak memory usage}
In \Tab{peak_memory}, we show the maximum RAM memory used for various algorithms in a single run. We can see that \THISWORK\ and PRESTO have comparable peak memory usage. Notice that the peak memory usage of \THISWORK\  does not grow with the motif's complexity or the number of samples. Because most of the memory is used to store the preprocessing weights.

\section{\Revision{Conclusion and Future Work}}
This paper presents \THISWORK---a general algorithm for counting motifs in temporal graphs that is both fast and accurate.
\THISWORK\ works for any motif with arbitrary size and shape.
The key idea behind \THISWORK\ is to estimate the motif counts based on an underlying spanning tree structure.
To improve estimation accuracy and achieve high performance, we show how to intelligently design a constrained algorithm and present the heuristics for choosing spanning trees.
\Revision{\THISWORK\ is designed under the assumption of total temporal orderings following temporal motif mining problems in prior works~\cite{mackey2018chronological}. A natural next step is to 
try to count motifs with partial order time constraints. In some special
cases, the spanning tree of \THISWORK\ suffices. 
Extending \THISWORK\ to fully support arbitrary partial orderings is an interesting direction for future work.}

\section{Acknowledgements}

Both OB and CS are supported by NSF grants CCF-1740850, CCF-2402572, and DMS-2023495.

\bibliographystyle{ACM-Reference-Format}
\bibliography{refs}


\end{document}